\documentstyle[12pt,epsfig,multirow]{article}

\newcommand{\be}{\begin{eqnarray}}
\newcommand{\ee}{\end{eqnarray}}

\def\nue{{\nu_e}}
\def\anue{{\bar\nu_e}}
\def\numu{{\nu_{\mu}}}
\def\anumu{{\bar\nu_{\mu}}}
\def\nutau{{\nu_{\tau}}}

\newcommand{\kl}{\mbox{KamLAND~}}

\newcommand{\ms}{\Delta m^2_{21}}
\newcommand{\ma}{\Delta m^2_{31}}

\newcommand{\sss}{\sin^2 \theta_{12}}

\newcommand{\stch}{\sin^2 2\theta_{13}}
\newcommand{\sa}{\sin^2 \theta_{23}}
\newcommand{\sta}{\sin^22 \theta_{23}}

\newcommand{\mat}{\Delta m^2_{31}{\mbox {(true)}}}

\newcommand{\stcht}{\sin^2 2\theta_{13}{\mbox {(true)}}}

\newcommand{\stat}{\sin^22 \theta_{23}{\mbox {(true)}}}

\newcommand{\sig}{$3\sigma$}
\newcommand{\bb}{$\beta$-beam~}
\newcommand{\bbf}{$\beta$-beam}
\def\ltap{\ \raisebox{-.4ex}{\rlap{$\sim$}} \raisebox{.4ex}{$<$}\ }
\def\gtap{\ \raisebox{-.4ex}{\rlap{$\sim$}} \raisebox{.4ex}{$>$}\ }

\begin{document}

\thispagestyle{empty}
\begin{flushright}
\texttt{hep-ph/0610333}\\
\texttt{HRI-P-06-10-001}\\
\texttt{CU-PHYSICS-17/2006}\\
\end{flushright}
\bigskip

\begin{center}
{\Large \bf Neutrino mass hierarchy and $\theta_{13}$ with
a magic baseline beta-beam experiment}

\vspace{.5in}

{\bf Sanjib Kumar Agarwalla$^{\star,\dagger,a}$, 
Sandhya Choubey$^{\star,b}$, Amitava Raychaudhuri$^{\star,\dagger,c}$}
\vskip .5cm
$^\star${\normalsize \it Harish-Chandra Research Institute,} \\
{\normalsize \it Chhatnag Road, Jhunsi, Allahabad  211019, India}\\
\vskip 0.4cm
$^\dagger${\normalsize \it Department of Physics, University of Calcutta,} \\ 
{\normalsize \it 92 Acharya Prafulla Chandra Road, Kolkata  700009, India}
\vskip 2cm

{\bf ABSTRACT}

\end{center}

We underscore the physics advantage of an
experiment  where  
neutrinos produced in a beta-beam facility at CERN 
are observed  
in a large magnetized iron calorimeter (ICAL) at 
the India-based Neutrino Observatory (INO). The CERN-INO
distance is close to the so-called ``magic" baseline which helps
evade some of the parameter degeneracies 
and allows for a better measurement of the neutrino mass
hierarchy and $\theta_{13}$. We expound the possibility of 
using radioactive $^8B$ and $^{8}Li$ as the source isotopes 
for the $\nue$ and $\anue$ beta-beam, respectively, and show that
very good sensitivity to both the mass hierarchy and $\theta_{13}$ 
is possible with a boost $\gamma$ in the  
250-500 ballpark.

\vskip 3cm

\noindent $^a$ email: sanjib@mri.ernet.in

\noindent $^b$ email: sandhya@mri.ernet.in 

\noindent $^c$ email: raychaud@mri.ernet.in

\newpage

\section{Introduction}

Spectacular results from a series of 
experiments involving solar \cite{solar},
atmospheric \cite{atm}, reactor \cite{kl},  
and accelerator \cite{k2k,minos} neutrinos have 
firmly established neutrino oscillations 
and heralded the precision era in
neutrino physics. 
Next generation 
experiments have been planned/proposed 
world-wide to further pin down the values of the  
oscillation parameters, $\ms$\footnote{We 
define $\Delta m_{ij}^2 = m_i^2 - m_j^2$.}
and $\theta_{12}$ for solar neutrinos
and $\ma$ and $\theta_{23}$ for atmospheric neutrinos. 
However, a vital task to be undertaken in the immediate future is
the determination of the hitherto unknown mixing angle
$\theta_{13}$\footnote{Currently, only an upper bound
($\sin^22\theta_{13} < 0.17$ at \sig{} \cite{chooz,limits}) exists.}.
Discovery of a non-zero value for $\theta_{13}$ would open up the
possibility of observing CP-violation in the lepton sector.
Another outstanding problem concerns the ordering of the neutrino
mass states, {\it aka}, the neutrino mass hierarchy.  If Nature
entertains a large enough mixing angle $\theta_{13}$,
it could be possible to ascertain $sgn(\ma)$ and hence the
neutrino mass hierarchy\footnote{The neutrino mass hierarchy is
termed ``normal'' (``inverted'') if $\ma$ is
positive (negative). Note that the terms normal and 
inverted ``hierarchy'' in this 
article refers to the neutrino mass ordering only, {\it i.e.},
whether $\nu_3$ is heavier or lighter than $\nu_2$,
and our discussions are valid 
even for a quasi-degenerate neutrino mass spectrum.}.

The neutrino mass hierarchy can be probed in terrestrial
experiments through matter effects 
\cite{msw1,msw2,magicfirst}.  Large
matter effects are known to exist in the oscillation probability
$\nue\rightarrow \numu$ ($P_{e\mu}$).  Since it also provides the
best way of determining the mixing angle $\theta_{13}$ and the CP
phase, $\delta_{CP}$, it has been popularly hailed as the
``golden channel'' \cite{golden}.  The $P_{e\mu}$ channel can be
studied in experiments which use an initial $\nue$ (or $\anue$)
beam and a detector which can efficiently see muons\footnote{This
is complementary to the standard accelerator beams where the
initial (anti)neutrino flux consists of $\numu$ (or $\anumu$) and
where the relevant channel is the
probability $P_{\mu e}$. Produced from
decay of accelerated pions, these conventional (anti)neutrino
beams suffer from an additional hurdle of an intrinsic $\nue$
($\anue$) contamination, which poses a serious problem of backgrounds.
A \bb is comprised of pure $\nue$
(or $\anue$).}.  An absolutely pure and intense $\nue$ (or
$\anue$) flux can be produced using   ``Beta Beams''
($\beta$-beam) \cite{zucc}.  There is a possibility of such a
facility coming up at the CERN accelerator complex and an
international R\&D effort is underway to check the feasibility of
this proposal \cite{iss}. The Tevatron at FNAL has also been
discussed as a potential accelerator for producing $\beta$-beams.
Most studies involving beta-beams have concentrated on using
megaton water \u{C}erenkov detectors for observing oscillations
through muons produced {\em via} $\numu$ (or $\anumu$).  The
reason is that these analyses consider low to medium values for
the acceleration of the $\beta$ unstable ions which create a
relatively low energy (anti)neutrino beam.  The detector
therefore should be capable of observing the resultant low energy
muons efficiently and unambiguously. Water \u{C}erenkov detectors
are particularly well suited for this. However, in contrast to
most of the previous work involving $\beta$-beams, which have
mainly shown the physics potential of these kinds of experiments
in determining $\theta_{13}$ and $\delta_{CP}$, in this paper we
are especially interested in deciphering the neutrino mass
hierarchy and $\theta_{13}$ simultaneously.  To that end, we
suggest a \bb with (anti)neutrino energies where we expect the
largest earth matter effects, that is, energies in the multi-GeV
regime.  These high energy $\numu$ (or $\anumu$) can be very
efficiently observed in large magnetized iron detectors, such as
the envisaged ICAL detector at the India-based Neutrino
Observatory \cite{ino}.
 
A serious complication with all long baseline experiments 
involving the golden $P_{e\mu}$ channel arises from  
degeneracies which manifest in three forms:
\begin{itemize}
\item the ($\theta_{13},\delta_{CP}$) intrinsic degeneracy 
\cite{intrinsic},
\item the ($sgn(\ma),\delta_{CP}$) degeneracy \cite{minadeg},
\item the ($\theta_{23},\pi/2-\theta_{23}$) degeneracy 
\cite{th23octant}.
\end{itemize}
This leads to an eight-fold degeneracy \cite{eight}, with several
spurious or ``clone'' solutions in addition to the true one and 
severely deteriorates the sensitivity of any experiment.  It has
been shown \cite{eight,magic} that the problem of clone solutions
due the first two types of degeneracies can be evaded by choosing
the baseline of the experiment equal to the characteristic
refraction length due to the matter inside earth
\cite{msw1,eight,magic,magic2}.  This special value goes by the
colloquial name ``magic baseline'' \cite{magic}. As we will
discuss in detail later, at this baseline the sensitivity to the
mass hierarchy and, more importantly, $\theta_{13}$, goes up
significantly \cite{magic,optimnufact}, while the sensitivity to
$\delta_{CP}$ is absent.

Interestingly, the CERN-INO distance of 7152 km happens to be
tantalizingly close to the magic baseline.  This large baseline
also enhances the matter effect and  requires  traversal through
denser regions of the earth. Thus, for neutrinos (antineutrinos)
with energies in the range 3-8 GeV sizable matter effects are
induced if the mass hierarchy is normal (inverted). A unique
aspect of this set-up is the possibility of observing
near-resonant matter effects in the $\nue\rightarrow \numu$
channel. In fact, to our knowledge, what we propose here is the
only experimental situation where near-resonant matter effects
can be effectively used in a long baseline experiment to study
the neutrino mass matrix.  We show in this paper that the
presence of this near-maximal earth matter effect not only
maximizes the sensitivity to the neutrino mass hierarchy, it also
gives the experiment an edge in the determination of the mixing
angle $\theta_{13}$. The increase in the probability $P_{e\mu}$
due to near-resonant matter effects, compensates for the fall in
the \bb flux due to the very long baseline, so that one can
achieve sensitivity to $\theta_{13}$ and mass hierarchy which is
comparable, even better, than most other proposed experimental
set-ups.  Therefore a \bb experiment with its source at CERN and
the detector at INO could emerge as a powerful tool for a
simultaneous determination of the neutrino mass hierarchy and
$\theta_{13}$.

In \cite{paper1} such an experimental set-up was considered for
the first time and the physics potential explored. It was
demonstrated that both the neutrino mass hierarchy as well as
$\theta_{13}$ may be probed through such a set-up.
The ions considered for the 
\bb in that work were the 
most commonly used 
$^{18}Ne$ (for $\nue$) and $^6He$ (for $\anue$). One 
requires very high values of the Lorentz boost for 
these ions ($\gamma \sim 10^3$) 
because 
the energy $E$ of the beam has to be in the 
few GeV range to enable detection in the ICAL 
detector, which is expected to have a threshold of 
about 1 GeV\footnote{Most studies on the physics 
reach of \bb experiments have energies in the few 
hundred MeV range.}.
Such high values of $\gamma$, although possible 
in principle, might turn out to be very difficult to realize. 
Very recently two other ions, $^8B$ (for $\nue$) and
$^8Li$ (for $\anue$),  have been projected as viable options for
the \bb source \cite{rubbia}.  The main advantage that these ions
offer  is their substantially higher end-point energy, $E_0$.
This allows one to access $E\sim$ few GeV very easily with medium
values of $\gamma$, that could be possible to achieve with either
the existing CERN technology, or with the projected upgrades.
The physics potential of these alternative ions 
to study standard neutrino oscillation parameters
in the context of a CERN to Gran Sasso \bb experiment 
has been expounded in \cite{doninialter}, while 
in \cite{rparity} they have been studied 
to test R-parity violating SUSY.

In this paper we will work with $^8B$ and $^8Li$ as 
the candidate ions
for the \bb source at CERN and ICAL at INO as the far detector. 
We perform a $\chi^2$ analysis of the 
projected data in a future CERN-INO 
\bb experiment and present our results for three plausible 
benchmark values of the Lorentz boost factor $\gamma$. 
We begin in section 2 with a brief description of the 
\bb produced at the source. In section 3 we 
discuss the oscillation probability and 
highlight the importance of the magic baseline and the near-resonant
matter effects in the $P_{e\mu}$ channel. Section 4 contains 
our results for the event rate expected in the ICAL detector.  In
section 5 we present the details of the statistical analysis
used. The sensitivity of the results on the mixing angle $\theta_{13}$
are presented in section 6, while section 7 focuses 
on the sensitivity to the neutrino mass hierarchy.  In section 8
we consider exploiting  the potential of muon charge
identification at ICAL at INO through a \bb running
simultaneously in both the $\nue$ and $\anue$ mode.  We end in
section 9 with our conclusions.

\section{The $\beta$-Beam Fluxes}

\begin{figure}[t]
\includegraphics[width=8.0cm, height=7.0cm]{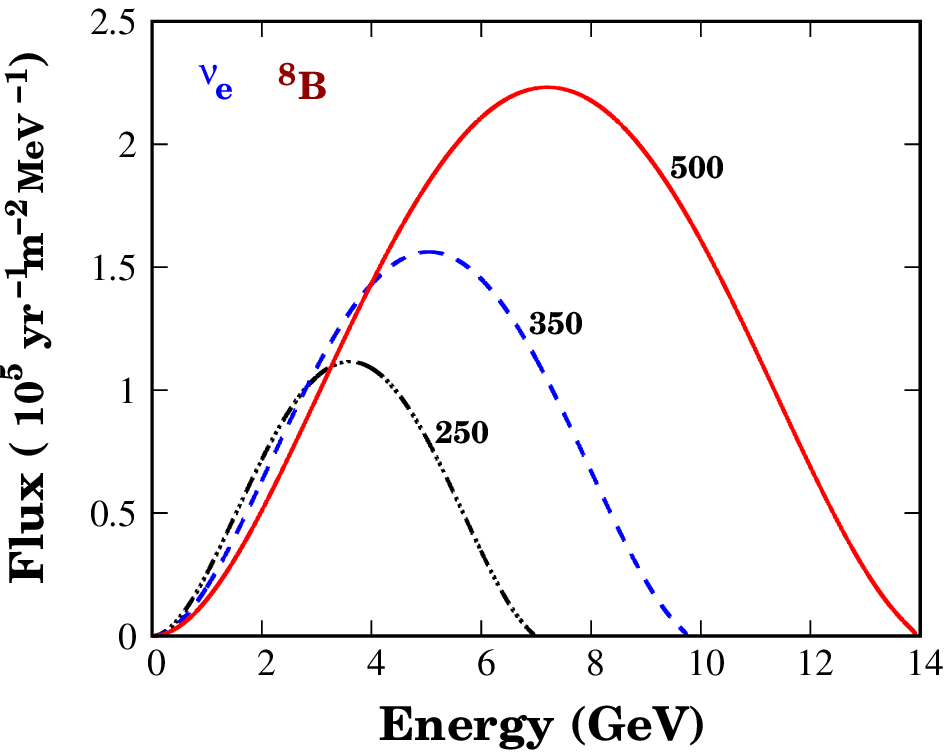}
\vglue -7.0cm \hglue 8.5cm
\includegraphics[width=8.0cm, height=7.0cm]{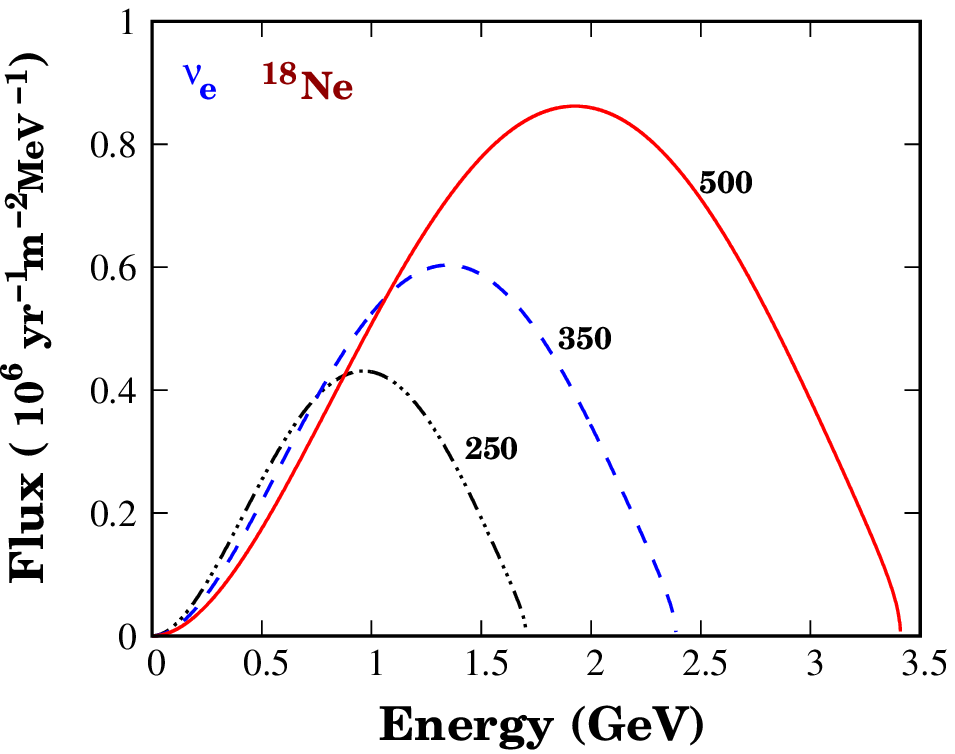}
\caption{\label{fig:nuflux}
The unoscillated $\nue$ spectrum at INO-ICAL 
from the $^8B$ (left-hand panel) and 
$^{18}Ne$ (right-hand panel) $\beta$-beam source at CERN, for 
three different choices of the boost factor $\gamma=250$ 
(black dot-dashed curve), 350 (blue dashed curve), and 500 (red solid curve).
}
\end{figure}
%
%
\begin{figure}[t]
\includegraphics[width=8.0cm, height=7.0cm]{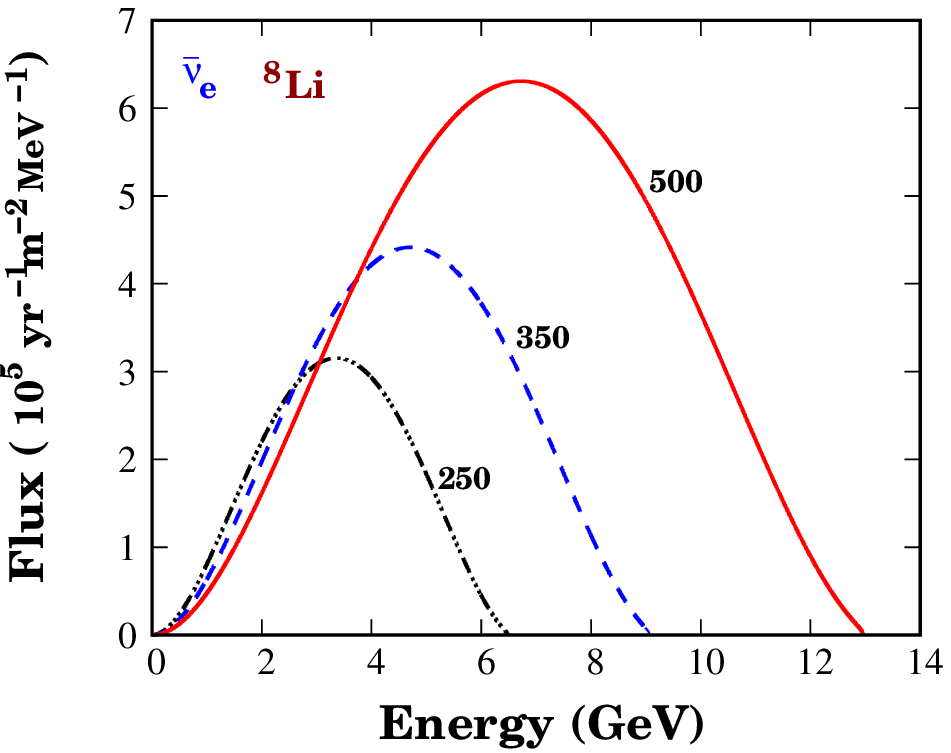}
\vglue -7.0cm \hglue 8.5cm
\includegraphics[width=8.0cm, height=7.0cm]{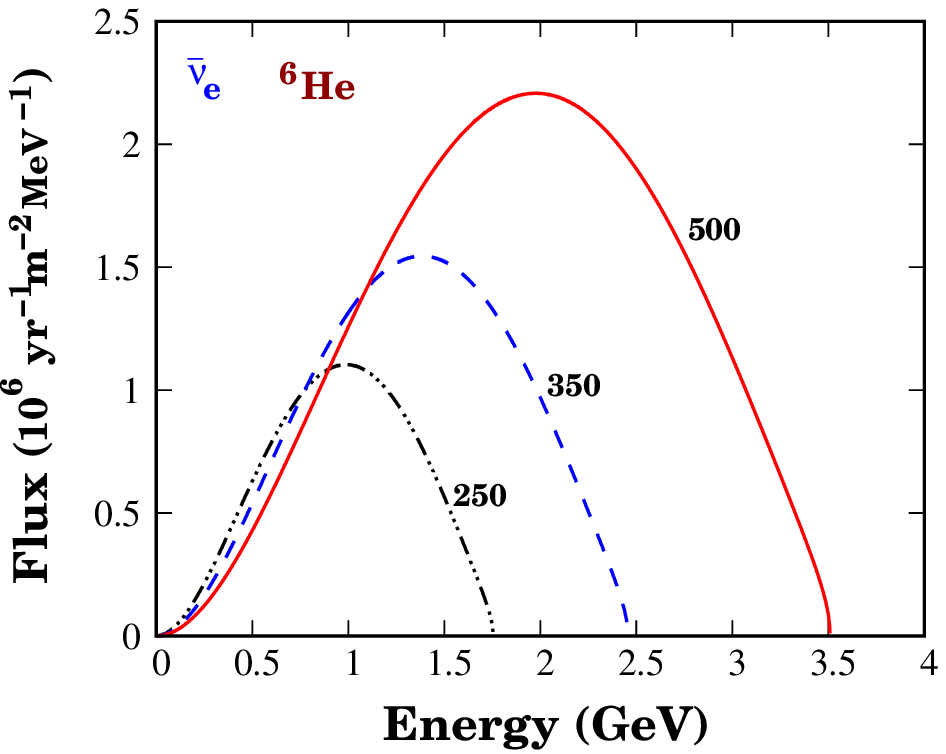}
\caption{\label{fig:anuflux}
Same as in Fig. \ref{fig:nuflux}, but for the $\anue$ spectrum from 
$^{8}Li$ (left-hand panel) and $^6He$ (right-hand panel) source.
}
\end{figure}

The concept of a \bb was proposed by Zucchelli 
\cite{zucc,volpe}. The idea is to produce a pure, intense, collimated
beam of $\nu_{e}$ or $\bar\nu_{e}$ through the beta
decay of highly accelerated and completely ionized 
radioactive ions circulating in a storage ring
\cite{jacques}. The resultant beam therefore comprises 
only of $\nue$ (or
$\anue$) and has definite characteristics, determined 
by the well known beta decay spectrum and the Lorentz boost factor.
The former feature results in very low backgrounds while the latter
ensures that there are essentially negligible flux systematics.
This proposal is being studied in depth and fully exploits
the existing CERN accelerator complex. The main future challenge lies in
manufacturing an intense proton driver 
and a hippodrome-shaped decay ring
which are at the core of this program. 

\subsection{The Beta Decay Spectrum}

In the rest frame of the radioactive ion the beta decay spectrum is
given by
\be
\frac{d^2 \Gamma^*}{d\Omega^* d E^*}
= \frac {1} {4\pi\,m_e^5 \,f 
\,\tau} (E_0 - E^*)E^* 
\sqrt{ (E_0-E^*)^2-m_e^2}
\label{eq:rspectra} ,
\ee
where $m_e$ is the electron mass, $E_0$ the electron 
total end-point energy, 
$E^*$ and $\tau$ are the neutrino
energy and lifetime of the decaying ion respectively 
in the latter's rest frame 
and
$f$ is a function of $m_e/E_0$.  
In the lab frame, 
the flux of the unoscillated \bb at the detector is given by
\be
\phi(E,\theta)
 =\frac{1}{4\pi L^2}\frac {N_\beta} {m_e^5 \,f} 
\frac{1}{\gamma(1-\beta \cos\theta)}
 (E_0 - E^*) E^{*2} \sqrt{ (E_0-E^*)^2-m_e^2},
\label{eq:flux}
\ee
where $L$ is the distance 
between the source and detector, $N_\beta$ are 
the number of useful decays in the storage ring 
per unit time, 
$\theta$ is the angle between 
the neutrino flight direction and the direction in which 
the ions are boosted\footnote{We work with the on-axis flux for which 
$\theta=0$.}
and $\gamma$ 
is the Lorentz boost such that
$E^* =  \gamma E(1-\beta \cos\theta)$, $E$ being the neutrino 
energy in the lab frame.
The maximum energy of neutrinos produced by a \bb with a Lorentz 
factor $\gamma$ is given by
\be
E_{\rm max}=\frac{(E_0-m_e)}{\gamma(1-\beta\cos\theta)}.
\label{eq:enumax}
\ee

\subsection{Candidate Ions for the \bb}

\begin{table}
\begin{center}
\begin{tabular}{|c|c|c|c|c|c|} \hline
   Ion & $\tau$ (s) &
$E_0$ (MeV)
   & $f$& Decay fraction & Beam \\
\hline
  $^{18} _{10}$Ne &   2.41 & 3.92&820.37&92.1\%& $\nu_{e}$    \\
  $^6 _2$He   &   1.17 & 4.02&934.53&100\% &$\bar\nu_{e}$    \\
\hline
 $^{8} _5$B& 1.11 & 14.43&600684.26&100\%&$\nu_{e}$    \\
 $^8 _3$Li& 1.20 &13.47 &425355.16& 100\% & $\bar\nu_{e}$    \\
\hline
\end{tabular}
\caption{\label{tab:ions}
Beta decay parameters: lifetime $\tau$, 
electron total end-point energy 
$E_0$, $f$-value
and decay fraction for various ions~\cite{beta}.}
\end{center}
\end{table}

From Eq. (\ref{eq:flux}) it is seen that the total flux and
energy of the \bb at the far detector depends mainly on the end
point energy $E_0$ of the beta decay ion and the Lorentz boost
factor $\gamma$. The flux increases as both $E_0$ and $\gamma$
increase. Larger $\gamma$ results in better collimation of the
impinging flux, thereby increasing the statistics.  The spectrum
also shifts to larger energies as $E_0$ and $\gamma$ increase.
The neutrino cross section in the detector
increases with energy, so for the same total flux a harder 
spectrum further enhances the statistics.  Since the flux falls as
$1/L^2$ as the source-detector distance increases, high values of
either $E_0$ or $\gamma$ or both are needed to have sizable
number of events at the far detector. $E_0$ is an intrinsic
property of the decaying ion while $\gamma$ is restricted by the
nature of the accelerators and the machine design. The other
desired properties which the ion should have include high
production yield, large decay fraction, and a life time that is
long enough to allow the ions to be adequately accelerated. It is
also easier to store larger number of lower-$Z$ isotopes in the
storage ring \cite{autin}.  In Table
\ref{tab:ions} we show the characteristic features of the four
different ions which have been discussed in the literature as
possible candidates for the $\beta$-beam. While $^{18}Ne$ and
$^8B$ are $\beta^+$ emitters (producing a $\nue$ beam), $^6He$
and $^8Li$ are $\beta^-$ emitters (producing a $\anue$ beam).
They all have comparable life times, conducive to the
requirements necessary for the \bbf, very high (or maximal) decay
fraction and very low $A/Z$ ratio.  Note that the
end-point energies, $E_0$, for $^8B$ and $^8Li$ are much larger
than those for  $^{18}Ne$ and $^6He$.

In this paper we are interested in the physics potential of 
an experimental set-up with a \bb at CERN and a large magnetized 
iron detector in India. For such a large baseline, one needs a
very high value of $\gamma$, both to cross the
detector energy threshold as well as to get reasonable statistics
in the detector. In particular, with $^6He$ (or $^{18}Ne$) as the
source ion, one needs $\gamma \gtap 1000$.  Such high values of
$\gamma$ can only be achieved by using the LHC itself. As noted
above, $E_0$ for the alternative options  
for the radioactive ion source ($^8B$ and $^8Li$) is higher by a factor of
about four.  This means that it should be possible to produce
high intensity, high energy beams with $^8B$ and $^8Li$ for a much
lower  boost factor\footnote{The loss in collimation is
not significant as we show later.}. We show in Fig.
\ref{fig:nuflux}  the $^8B$ (left-hand
panel) and $^{18}Ne$ (right-hand panel) \bb flux  
expected at INO as a
function of the $\nue$ energy, for three
different benchmark values of $\gamma$.  Fig.  \ref{fig:anuflux}
shows the corresponding spectra for the $^8Li$ (left-hand panel)
and $^{6}He$ (right-hand panel) $\anue$ $\beta$-beam.  Note that
even though apparently it might seem from the figures that the
$\nue$ ($\anue$) flux is larger for $^{18}Ne$
($^6He$), in reality, for a given $\gamma$, the total flux is
given by the area under the respective curves.    One can easily
check that for a fixed $\gamma$ this is same for both the ions as
expected, since we have assumed equal number of decays for both
$^{18}Ne$ and $^8B$ for $\nue$ ($^6He$ and $^8Li$ for
$\anue$)\footnote{Of course, the on-axis flux  increases with
$\gamma$ because of better collimation of the beam.}.  Also
note that even though the total $\nue$ ($\anue$) flux remains the
same for both the ions for a given $\gamma$, the number of events
produced in the detector is much enhanced for the $^8B$ ($^8Li$)
beam since the energy of the beam is larger and the charged
current cross sections increase with the neutrino energy.
Studies have shown that it is possible to accelerate $^6He$ to
$\gamma \ltap 250$ with the existing facilities at CERN, while
$\gamma=250-600$ should be accessible with the ``Super-SPS'', an
upgraded version of the SPS with super-conducting magnets
\cite{ssps,lhcupgrades}.  
The Tevatron at FNAL could in principle also be
used to produce a \bb with $\gamma < 600$.

In the low $\gamma$ design of
beta-beams 
$2.9\times 10^{18}$ $^6He$ and $1.1\times
10^{18}$ 
$^{18}Ne$ useful decays per year should be possible to achieve.
Earlier,
only these ions were considered because it was believed that
$^8B$ could not be produced with the standard ISOLDE techniques.
Since most exercises focused on observation of CP-violation, it
was necessary to have both $\nue$ and $\anue$ beams with similar
spectra, so $^8Li$ (though considered in \cite{autin}) was also
generally ignored.  Interest in both these ions have been
rekindled recently \cite{rubbia,doninialter}, as it appears that
having intense $^8B$ and $^8Li$ fluxes should be feasible using
the ionization cooling technique \cite{rubbia}. In what follows,
we will vary the value of $\gamma$ to test the physics potential
of the CERN \bb INO-ICAL set-up and, following the current
practice, assume that it is possible to
get $2.9\times 10^{18}$ useful decays per year for $^8Li$ and $1.1\times
10^{18}$ for $^8B$ for all values of $\gamma$.

\section{Oscillation Phenomenology -- The Magic Baseline}
\begin{table}[t]
\begin{center}
\begin{tabular}{|l|}
\hline
\\[-.5mm]
$|\Delta m^2_{31}| = 2.5 \times 10^{-3} \ {\rm eV}^2$ \\[2mm]
$\sin^2 2 \theta_{23} = 1.0$ \\[2mm]
$\Delta m^2_{21} = 8.0 \times 10^{-5} \ {\rm eV}^2$ \\[2mm]
$\sin^2\theta_{12} = 0.31$ \\[2mm]
$\delta_{CP} = 0$ \\[2mm]
\hline
\end{tabular}
\caption{\label{tab:true}
Chosen benchmark values of oscillation parameters, except
$\stch$.}
\end{center}
\end{table}

\begin{figure}[t]
\begin{center}
\includegraphics[width=16.0cm, height=9.5cm]{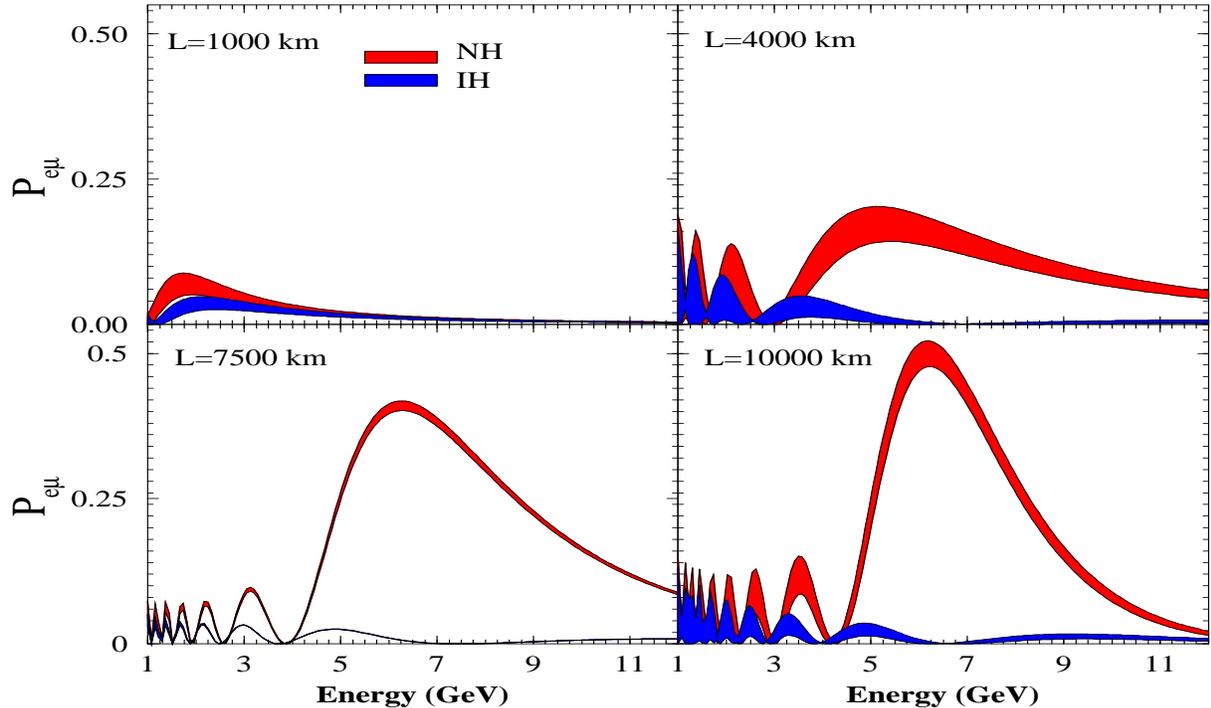}
\caption{\label{fig:p12Efixedcp}
The transition probability $P_{e\mu}$ as a function of $E$ 
for four values of the baseline $L$. 
The band reflects the effect of the 
unknown $\delta_{CP}$. The dark (red) shaded band is for 
the normal hierarchy (NH) while the light (cyan) shaded 
band is for the 
inverted hierarchy (IH). We have taken $\stch=0.1$ and 
for all other oscillation parameters we assume the 
benchmark values given in Table \ref{tab:true}.}
\end{center}
\end{figure}
%
\begin{figure}[t]
\begin{center}
\includegraphics[width=16.0cm, height=9.5cm]{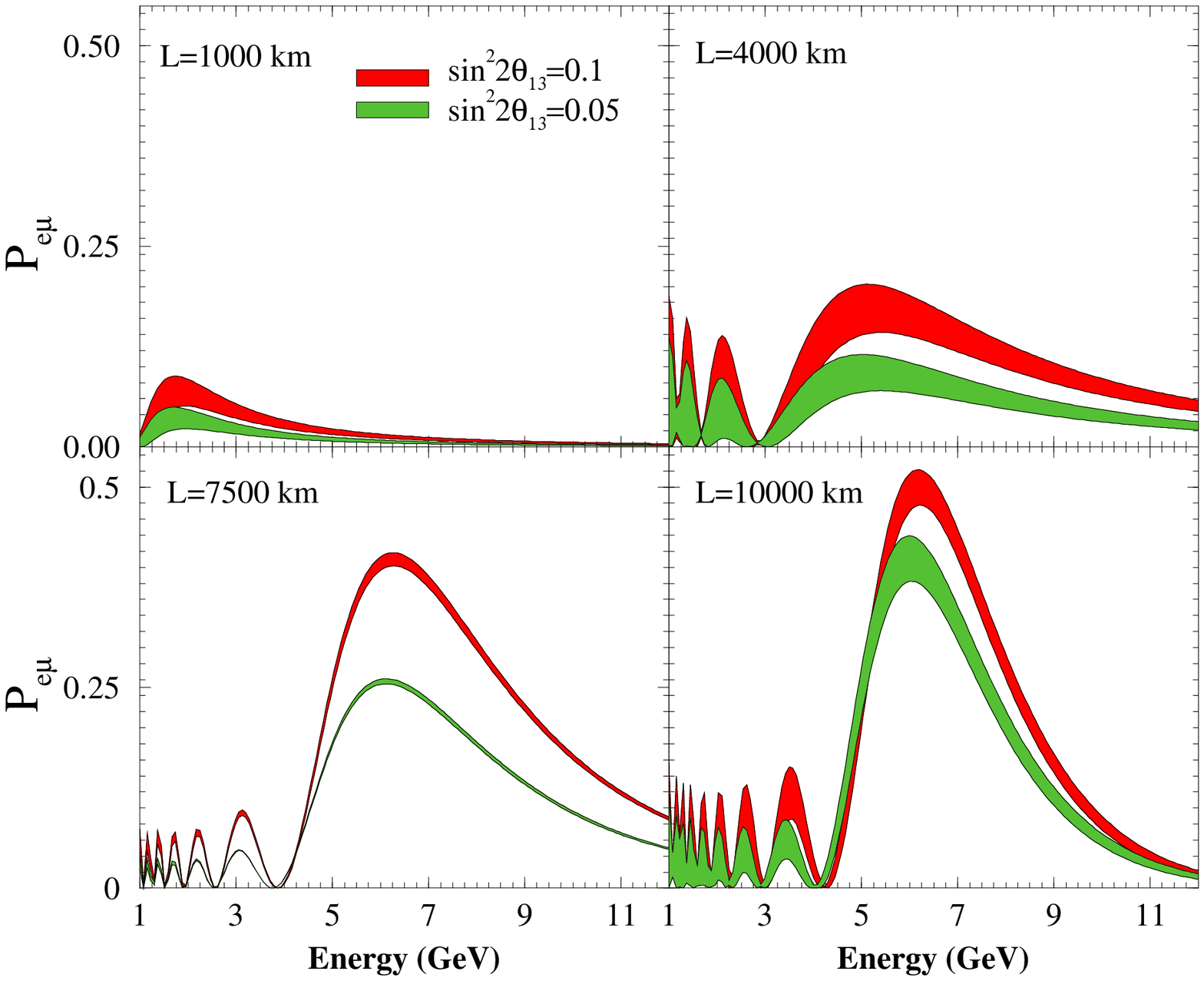}
\caption{\label{fig:p12Efixedt13}
The dark (red) shaded band is the same as in Fig. \ref{fig:p12Efixedcp}.
The light (green) shaded band shows the corresponding $P_{e\mu}$ 
for $\stch=0.05$. Values of all the other oscillation parameters 
are same as in Fig. \ref{fig:p12Efixedcp} and the hierarchy is assumed 
to be normal.}
\end{center}
\end{figure}
%

In this section we will very briefly review some issues related 
to the neutrino oscillation probability. We consider the 
standard form of the neutrino mixing matrix
\be 
U = \pmatrix
{c_{12} c_{13} & s_{12} c_{13} & s_{13} e^{-i \delta_{CP}} 
\cr
-s_{12} c_{23} - c_{12} s_{23} s_{13} e^{i \delta_{CP}} 
& c_{12} c_{23} - s_{12} s_{23} s_{13} e^{i \delta_{CP}} 
& s_{23} c_{13} \cr
s_{12} s_{23} - c_{12} c_{23} s_{13} e^{i \delta_{CP}} 
& 
-c_{12} s_{23} - s_{12} c_{23} s_{13} e^{i \delta_{CP}} 
& c_{23} c_{13}\cr
} 
,
\label{eq:upmns} 
\ee
where $c_{ij}=\cos\theta_{ij}$, $s_{ij}=\sin\theta_{ij}$
and we ignore the Majorana phases.
From the analyses of the current solar, atmospheric, 
reactor, and accelerator data set one has the following \sig{}
constraints on the oscillation parameters \cite{limits}:
\be
7.2\times 10^{-5} {\rm eV}^2 < \ms < 9.2\times 10^{-5} {\rm eV}^2 \;
{\rm and}\;
0.25 < \sss < 0.39 \;,
\ee
\be
2.0\times 10^{-3} {\rm eV}^2 < |\ma| < 3.2\times 10^{-3} {\rm eV}^2 \;
{\rm and}\;
\sin^22\theta_{23} > 0.9 \;,
\ee
\be
\stch < 0.17 \;,
\label{eq:sch}
\ee
Throughout this paper we show all our results assuming certain
true values for the oscillation parameters $|\ma|$, $\sa$, 
$\ms$, $\sss$ and $\delta_{CP}$ given in Table \ref{tab:true}.
We will let $\stch$ take all possible true values (which we will 
call $\stcht$) and present our result as a function of $\stcht$.

\noindent
Since both $\stch$ and $\alpha \equiv \ms/\ma$ are very 
small parameters, as can be seen from the current constraints,
we can expand the so-called ``golden channel''
probability in the constant matter density approximation 
as\footnote{This particular low order expansion of the 
transition probability is valid only in the range
of $L$ and $E$ 
where the resonance 
condition $\hat{A}=1$ is never reached. For the $E$ and $L$ 
range that we consider in this paper, this condition is satisfied 
and Eq. (\ref{eq:pemu}) fails, as we discuss later in this section.
Nonetheless, we present this expression to illustrate 
the effect of the magic baseline.} \cite{golden,freund}
\be
P_{e\mu} &\simeq& 
 \sin^2\theta_{23} \sin^22\theta_{13}
\frac{\sin^2[(1-\hat{A})\Delta]}{(1-\hat{A})^2}\nonumber \\
&+& \alpha \sin2\theta_{13} \sin2\theta_{12} \sin2\theta_{23} 
\sin\delta_{CP} \sin(\Delta) \frac{\sin(\hat{A}\Delta)}{\hat{A}}
\frac{\sin[(1-\hat{A})\Delta]}{(1-\hat{A})} \nonumber \\
&+& \alpha \sin2\theta_{13} \sin2\theta_{12} \sin2\theta_{23} 
\cos\delta_{CP} \cos(\Delta) \frac{\sin(\hat{A}\Delta)}{\hat{A}}
\frac{\sin[(1-\hat{A})\Delta]}{(1-\hat{A})} \nonumber \\
&+& \alpha^2 \cos^2\theta_{23} \sin^22\theta_{12} 
\frac{\sin^2(\hat{A}\Delta)}{{\hat{A}}^2}
,
\label{eq:pemu}
\ee
where 
\be
\Delta\equiv \frac{\ma L}{4E},
\ee 
\be
\hat{A} \equiv \frac{\pm A}{\ma},
\label{eq:matt}
\ee
$A= 2\sqrt{2}G_FN_eE$ being the matter potential, where $N_e$ is the 
electron number density inside the earth
and we have expanded the probability 
in $\alpha$ and $\sin2\theta_{13}$, keeping only 
up to second order terms in both these parameters. 
The + ( $-$ ) sign 
in Eq. (\ref{eq:matt})
refers to neutrinos (antineutrinos). The first term of 
Eq. (\ref{eq:pemu}) can be used to extract information about the 
value of $\theta_{13}$. This is also the term which has the largest 
earth effect and this effect of matter can be used to 
determine the sign of $\ma$.  The second (CP violating) 
and third (CP conserving) terms depend on 
the CP phase $\delta_{CP}$ and 
can be used to find it. The last term is independent 
of both $\theta_{13}$ and $\delta_{CP}$ and depends mainly 
on the solar parameters $\ms$ and $\theta_{12}$. In very long
baseline experiments, this term has sizable matter effects and 
if the true value of $\theta_{13}$ turns out to be zero (or nearly zero),
this would be the only surviving term in $P_{e\mu}$ 
which could still be used to 
study matter enhanced oscillations \cite{d21msw}.

Despite its advantage in determining the most interesting 
oscillation parameters, this channel is, however, rife
with the problem of ``degeneracies'' -- 
the $\theta_{13}$-$\delta_{CP}$ intrinsic degeneracy \cite{intrinsic},
the $sgn(\ma)$
degeneracy \cite{minadeg}
and the octant of $\theta_{23}$ degeneracy \cite{th23octant} --  
leading to an overall 
eight-fold degeneracy in the parameter values \cite{eight}. 
Various schemes to tackle some or all of these problems 
have been studied in great detail in the literature. 
These can be broadly categorized into ones where data from 
experiments at different energies and 
different baselines are combined together 
\cite{intrinsic,t2kplusnova,t2kk-I,t2kk-II,novadeg,differentLnE},
where data from 
accelerator experiments observing different 
oscillation channels are considered 
together \cite{silver,Bueno:2001jd},
and where data from atmospheric \cite{t2katm,cernmemphys} or 
reactor experiments \cite{reactor} are combined with 
the golden channel measurements in an accelerator 
experiment.

A particularly interesting scenario arises when the condition
\be
\sin(\hat{A}\Delta)=0
\label{eq:condmagic}
\ee
is satisfied. In such an event, the last three terms in Eq.
(\ref{eq:pemu}) drop out and   
the $P_{e\mu}$ channel enables a clean determination 
of $\theta_{13}$ and $sgn(\ma)$. 

Since 
$\hat{A}\Delta = \pm A L/4E$ by definition, the condition 
(\ref{eq:condmagic}) reduces to 
$\rho L = \sqrt{2}\pi/G_F Y_e$, 
where $Y_e$ is the electron fraction inside the earth. This 
gives
\be
\frac{\rho}{[{\rm g/cc}]}\frac{L}{[km]} \simeq 32725~,
\ee
which for the PREM \cite{prem} density profile of the earth 
is satisfied for the ``magic baseline'' \cite{magic}
\be
L_{\rm magic} \simeq 7690 ~{\rm km}.
\ee
We refer the 
reader to \cite{magic2} 
for a very recent and enlightening discussion on the 
physical meaning of the magic baseline.

The CERN-INO
distance corresponds to $L=7152$ km, which is tantalizingly 
close to the magic baseline. We therefore expect that the
\bb experiment we consider here should give 
an essentially degeneracy-free measurement of 
both $\theta_{13}$ and $sgn(\ma)$. 

The large CERN-INO baseline, of course, results in very significant 
earth matter effects in the $P_{e\mu}$ channel. In fact, 
for the baseline of 7152 km, the average 
earth matter density calculated using the PREM earth 
matter density profile is $\rho_{av}=4.13$ gm/cc, for which 
the resonance energy
\be
E_{res} &\equiv& {|\ma| \cos 2\theta_{13} \over
2\sqrt{2} G_F N_e}~\\
&=& 6.1~{\rm GeV}~,
\label{eq:eres}
\ee
for $|\ma|=2.5\times 10^{-3}$ eV$^2$ and $\stch=0.1$. 
One can check (cf. Figs \ref{fig:nuflux}
and \ref{fig:anuflux})
that this is roughly in 
the ballpark where we expect the maximum \bb flux for $^8B$ and
$^8Li$ ions with the benchmark $\gamma$ values used in this paper.

Note that the low order expansion of the probability $P_{e\mu}$ 
given by Eq. (\ref{eq:pemu}) is valid only for values of 
$E$ and earth matter density $\rho$ (and hence $L$) 
where flavor oscillations are far from resonance, {\it i.e.}, 
$ \hat{A}\ll 1$. In the limit $\hat{A} \sim 1$, 
one can check that even 
though the analytic expression for $P_{e\mu}$ 
given by Eq. (\ref{eq:pemu}) remains finite,
the resultant probability obtained is 
incorrect \cite{Takamura:2005df}. 
We reiterate that Eq. (\ref{eq:pemu}) was presented only 
in order to elucidate the 
importance of the magic baseline. For all the numerical 
results presented in this paper, we calculate the exact three 
generation oscillation probability using the 
realistic PREM \cite{prem} profile for the earth matter density.

The exact neutrino transition probability 
using the PREM density profile is 
given in Fig. \ref{fig:p12Efixedcp} as a function of the 
neutrino energy, for four different baselines 
(four panels). 
We allow $\delta_{CP}$ to take on 
all possible values between 0-360 degrees and 
the resultant probability is shown as a band, with the  thickness of 
the band reflecting the effect of $\delta_{CP}$ on $P_{e\mu}$.
The figure is drawn assuming the 
benchmark values of the oscillation parameters 
given in Table \ref{tab:true}
and $\stch=0.1$. 
We show the probability for both 
the normal hierarchy (dark band) as well as the inverted 
hierarchy (light band). As discussed in detail above, for 
$L=7500$ km, which is close to the magic baseline, the 
effect of the CP phase is seen to be almost negligible, 
while for 
all other cases the 
impact of $\delta_{CP}$ on $P_{e\mu}$ is seen to be 
appreciable. In fact, for $L=1000$ km, the probability 
corresponding to the normal and inverted hierarchy become almost 
indistinguishable 
due to the uncertainty arising from the unknown value 
of $\delta_{CP}$. 
As the baseline is increased, earth matter 
density increases, enhancing the impact of matter effects. 
The probability for normal hierarchy is hugely enhanced for 
the neutrinos, while for the inverted hierarchy matter effects 
do not bring any significant change. 
This difference in 
the predicted probability, evident 
in the panels corresponding to $L=4000$, 7500 and 10000 km
of Fig. \ref{fig:p12Efixedcp}, can be used to  
determine the neutrino mass hierarchy. 

In Fig. \ref{fig:p12Efixedt13} we display the dependence of the
neutrino probability $P_{e\mu}$ on the mixing angle $\theta_{13}$
for four different baselines. The dark bands, as in Fig.
\ref{fig:p12Efixedcp}, are for normal hierarchy and $\stch=0.1$
with full variation of $\delta_{CP}$, while the light bands are
for normal hierarchy and $\stch=0.05$.  The impact of matter
effect in increasing the $\theta_{13}$ sensitivity of a given
experimental set-up is evident from the figure. The $\theta_{13}$
sensitivity for $L=1000$ km can be seen to be much weaker than
for the other cases, since matter effects are smaller.
However, the most striking feature seen in Fig. 
\ref{fig:p12Efixedt13} is the effect of the 
magic baseline in enhancing the sensitivity of the 
experiment to $\theta_{13}$. The figure
clearly shows that the difference in 
the predicted $P_{e\mu}$ for the two values of $\stch$
is largest for $L=7500$ km (since effect of 
$\delta_{CP}$ is the least) and thus an experiment at this 
baseline is most suitable for probing $\theta_{13}$.
Figs. \ref{fig:p12Efixedcp} and \ref{fig:p12Efixedt13}
therefore reinforce our choice of the near-magic baseline
as one of the best options for determining the neutrino mass hierarchy
and $\theta_{13}$, since both these parameters
are directly related to large matter effects and  
the uncertainty of $\delta_{CP}$ 
could prove to be a hindrance in their measurement 
at non-magic baselines.

\section{Event Rates in INO-ICAL}

\subsection{The ICAL Detector at INO}

The proposed large magnetized iron calorimeter  at 
the India-based Neutrino Observatory  \cite{ino}
is planned 
to have a total mass of 50 kton at startup, which might be 
later upgraded to 100 kton. 
The INO facility is 
expected to come up at PUSHEP (lat. North 11.5$^\circ$,
long. East 76.6$^\circ$), situated close to Bangalore
in southern India. This constitutes a baseline of 
7152 km from CERN. The ICAL detector will have a 
modular structure with a total lateral size of 
$48{\rm m} \times 16 {\rm m}$, divided into three modules 
of $ 16{\rm m}\times  16 {\rm m}$ each. Each of these modules 
will have 140 horizontal  
layers of $\sim 6$ cm thick iron plates, separated from  
each other by a gap of $\sim 2.5$ cm to hold the
active detector material, giving a total 
height of 12 m for the full detector. The active 
detector elements will be resistive plate 
chambers (RPC), made from either glass or Bakelite and will be 
filled with a suitable gas mixture, which will be recycled with 
approximately one volume change per day. An external magnetic 
field of $\sim 1.3$ Tesla would be applied over the entire 
detector. The detector will be surrounded by an external layer of 
scintillator or proportional gas counters which will act both
as veto to identify external muon backgrounds as well as to identify
partially contained events.

According to the detector simulation performed by the INO
collaboration, the detector energy threshold for $\mu^\pm$ is
expected to be around $\sim 1$ GeV and charge identification
efficiency will be about 95\%. In what follows, we will present
our numerical results assuming an energy threshold of 1.5 GeV,
detector charge identification efficiency as 95\% and unless
stated otherwise, detection efficiency as 60\% for $\mu^\pm$ (cf.
Table \ref{tab:detector}). The 60\% detection efficiency we have
used is an extremely conservative estimate and it should be 
possible to increase it to at least 80\%.
We shall therefore also present our main results for this 
more optimistic estimate for the detector efficiency. However, 
unless it is mentioned otherwise, all our results 
correspond to the detector characteristics given in 
Table \ref{tab:detector}.
We have explicitly checked that our results remain unaffected 
if the energy threshold is raised to 2 GeV for the entire 
range of assumed Lorentz boost factor $\gamma=250-500$. 
For $\gamma > 350$ the threshold can be even 3 GeV,
while for $\gamma > 500$ one can work with an energy threshold 
of 4 GeV, without changing the 
final results. This can be seen from  
Figs. \ref{fig:nuflux} and \ref{fig:anuflux};  
for $\gamma=350(500)$, the majority of neutrinos arriving 
at INO-ICAL would have $E>3(4)$ GeV.

\begin{table}[h]
\begin{center}
\begin{tabular}{|l|c|}
\hline
&\\[-0.5mm]
Total Mass & 50 kton  \\[2mm]
Energy threshold & 1.5 GeV \\[2mm]
Detection Efficiency ($\epsilon$) & 60\% \\[2mm]
Charge Identification Efficiency ($f_{ID}$)& 95\%\\[2mm]
\hline
\end{tabular}
\caption{\label{tab:detector}
Detector characteristics used in the simulations.
}
\end{center}
\end{table}

\subsection{Oscillation Signal}

The total number of $\mu^-(\mu^+)$ events collected in INO-ICAL 
from a $\nue$ (or $\anue$) 
\bb exposure over a period of $T$ years is 
given by,
\be
N_{\mu} = T\, n_n\, f_{ID}\,\epsilon~  \int_0^{E^{\rm max}} dE 
\int_{E_A^{th}}dE_A \,\phi(E) \,\sigma_\numu(E) \,R(E,E_A)\, P_{e\mu}(E)
\label{eq:rates}
\ee
where $n_n$ are the number of target nucleons, $f_{ID}$ is the 
charge identification efficiency, $\epsilon$ is the detection 
efficiency, 
$\phi(E)$ (given by Eq. (\ref{eq:flux})) is the  
\bb flux at INO-ICAL in units of m$^{-2}$year$^{-1}$GeV$^{-1}$, 
$\sigma_\numu$ is the detection cross section for $\numu$ 
in units of m$^{2}$, 
$E$ is the true energy of the incoming neutrino
and $E_A$ is the measured energy\footnote{The ``measured'' 
energy of the neutrino can be reconstructed 
from the total measured energy of the muon and the 
accompanying hadron shower.
}$^,$\footnote{The 
integration over the reconstructed (measured) neutrino energy is done 
from the threshold energy $E_A^{th}$ (taken as 1.5 GeV in 
this paper) to infinity.},
$R(E,E_A)$ is the detector 
energy resolution function\footnote{
We assume a Gaussian resolution function with $\sigma=0.15E$.},
$E^{\rm max}$ is the maximum energy of the neutrinos
for a given Lorentz factor $\gamma$ (given by Eq. (\ref{eq:enumax}))
and $P_{e\mu}$ is the $\nue\rightarrow \numu$ 
oscillation probability.  
The expression for the 
$\mu^+$ signal in the detector from a $\anue$ \bb flux 
is given by replacing $\phi_\nue$ by $\phi_\anue$, $\sigma_\numu$ 
by $\sigma_\anumu$ and $P_{e\mu}$ by $P_{\bar e{\bar\mu}}$.
For the neutrino-nucleon interaction we consider quasi-elastic 
scattering, single-pion production  and deep inelastic 
scattering and use the cross sections given in the Globes package 
\cite{globes} which are taken from \cite{Messier:1999kj,Paschos:2001np}.

\begin{figure}
\includegraphics[width=8.0cm, height=7.0cm]{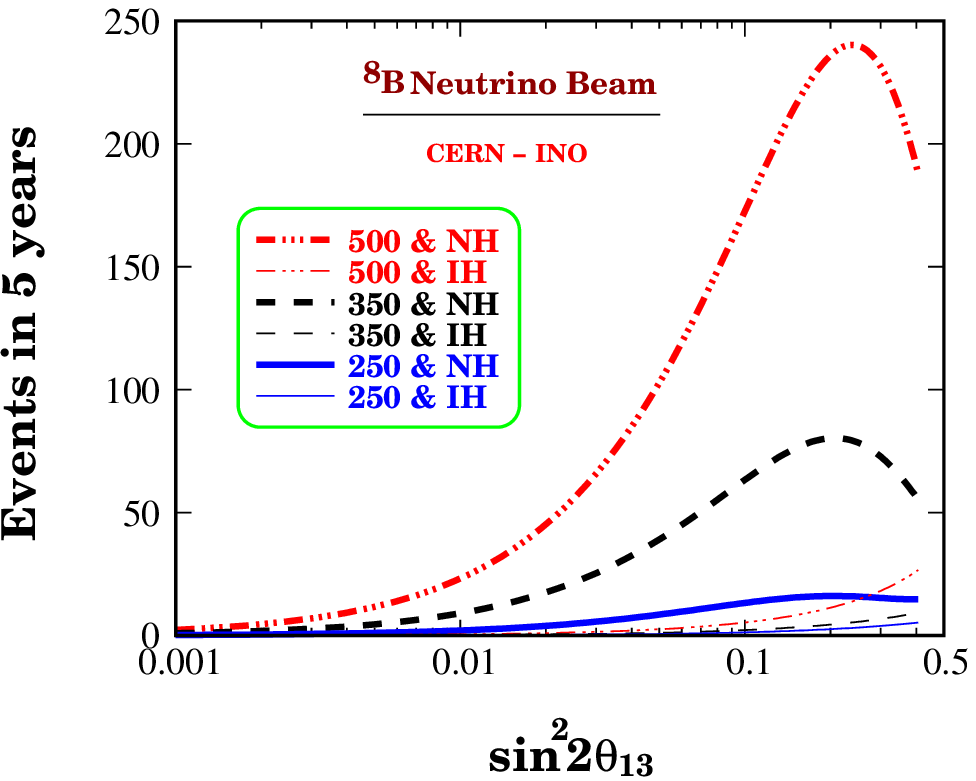}
\vglue -7.0cm \hglue 8.5cm
\includegraphics[width=8.0cm, height=7.0cm]{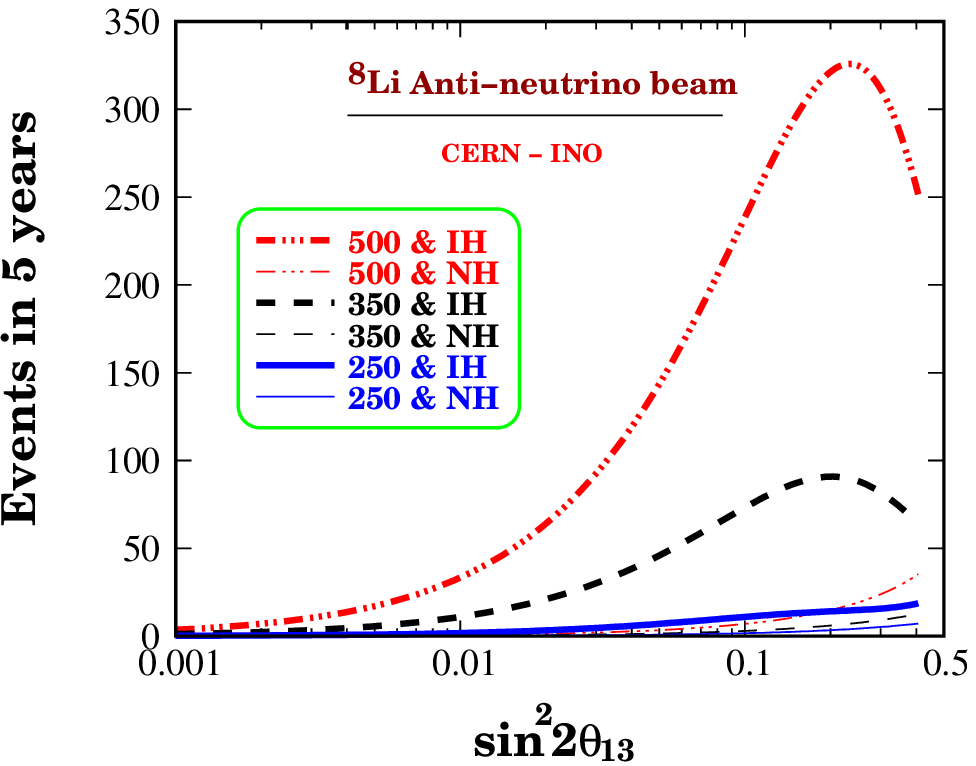}
\caption{\label{fig:rates}
The expected number of events in 
5 years running time, as a function of $\stch$. The value of $\gamma$ and 
the hierarchy chosen corresponding to each curve is shown in the 
figure legend.}
\end{figure}

\begin{figure}
\includegraphics[width=8.0cm, height=7.0cm]{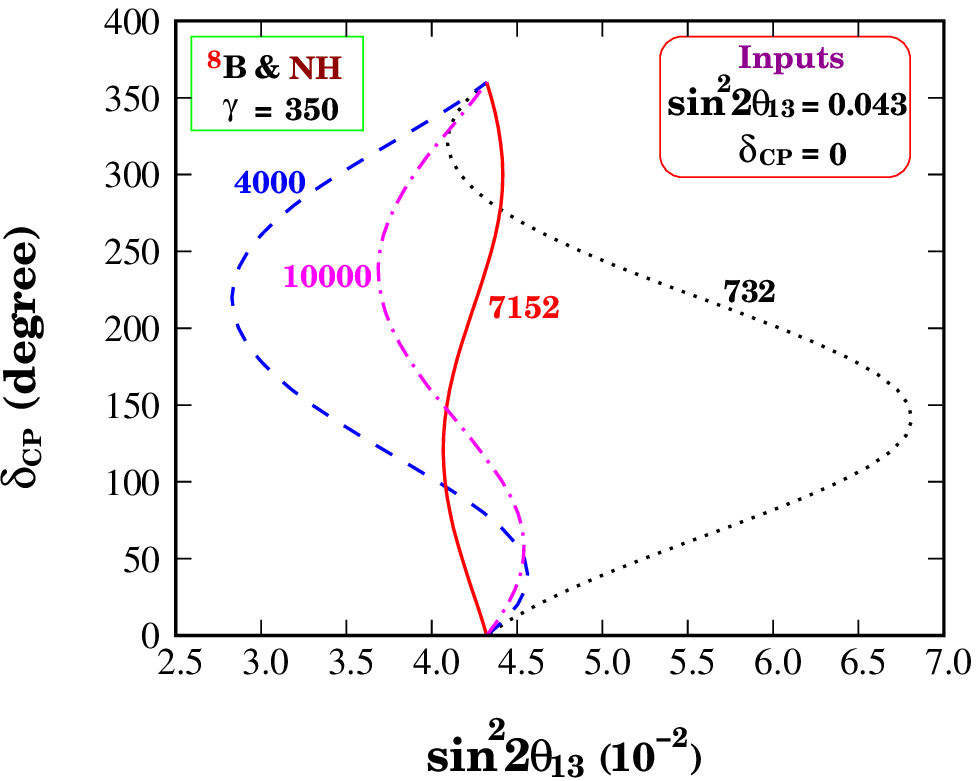}
\vglue -7.0cm \hglue 8.5cm
\includegraphics[width=8.0cm, height=7.0cm]{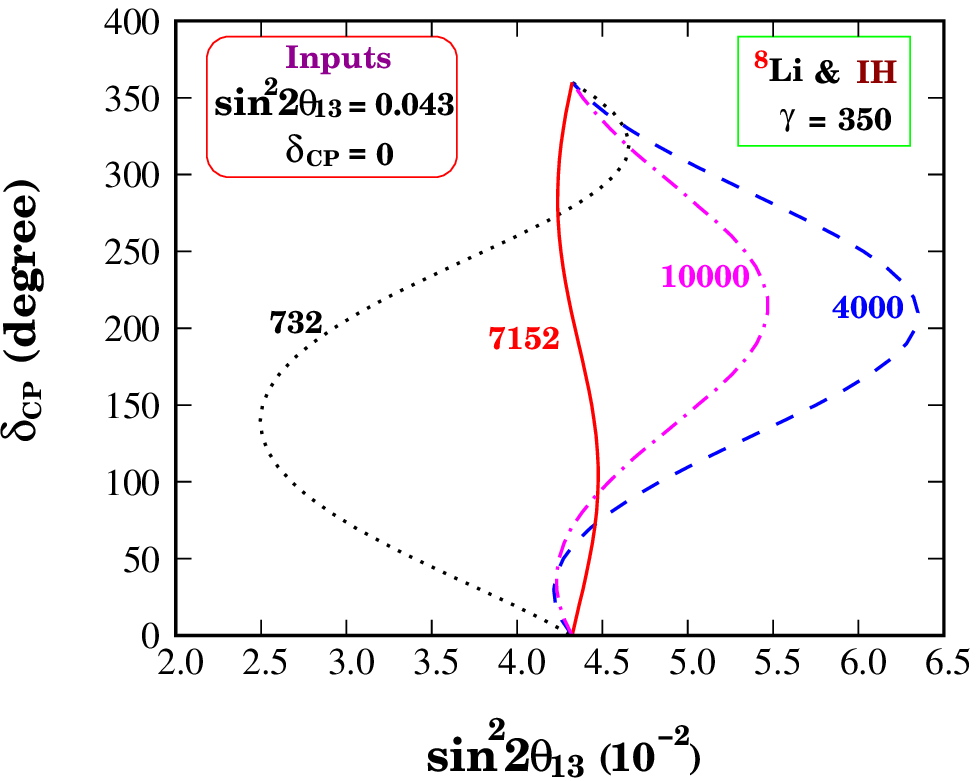}
\caption{\label{fig:deg}
The iso-event curves in the $\delta_{CP}$-$\stch$ plane
for four baselines are shown in the figure. The true 
values of $\delta_{CP}$ and $\stch$ are assumed to 
be $0^\circ$ and $0.043$ respectively. 
Left- (right-) hand panel
is for the $\nue$ ($\anue$) $\beta$-beam. The assumed 
hierarchy is mentioned in the figure.
}
\end{figure}

We show in Fig. \ref{fig:rates} the number of events expected 
in INO-ICAL
from 5 years exposure of an $^8B$ (for $\nue$, shown 
in the left-hand panel) or $^8Li$ 
(for $\anue$, shown in the right-hand panel) 
\bb from CERN. The expected number of events is presented as a
function of $\stch$ for both normal (NH) and inverted (IH) 
hierarchy for three benchmark values of the boost factor $\gamma$. 
Large resonant matter effects in the neutrino channel for 
normal hierarchy drives the number of expected events to very 
large values, compared to what would be expected for 
inverted hierarchy. 
Similarly, in the antineutrino channel 
we have resonant matter effects for 
inverted hierarchy and the number of predicted events is many times 
larger than for normal hierarchy\footnote{In fact, matter effects 
are seen to suppress the oscillation probability for 
neutrinos (antineutrinos) compared to that in vacuum when the
hierarchy is inverted (normal).  }.  This difference in the
number of events is seen to increase with $\stch$ up to a certain
(large) value, beyond which it starts to decrease.
This limiting value of $\stch$ at which the number of events in
the neutrino (antineutrino) channel peaks for the normal
(inverted) hierarchy comes from an interplay of two effects which
we now discuss. In the approximation where $\ms$ can be
neglected, the probability $P_{e\mu}$ would be largest if the
mixing angle in matter ($\sin^22\theta_{13}^M$) and the mass
squared difference driven oscillatory term in matter
($\sin^2[(\ma)^ML/4E]$) are maximum simultaneously.  In other
words, largest matter effects come when the resonance energy
$E_{res}$ 
is almost
equal to the energy at which the oscillatory factor goes to one
\cite{anuls:2001zn,gandhi1,gandhi2}.
For a given 
$L$ (and hence matter density, $\rho$), one can calculate the 
value of $\stch$ for which this condition can be 
satisfied and it is given by \cite{gandhi1,gandhi2}
\be
\tan2\theta_{13} = \frac{16.18\times 10^3}{\rho{\rm(gm/cc)} L{\rm(km)}}
\ee
For $L=7152$ km, it turns out that 
this value is $\stch\simeq 0.23$. Therefore, matter effects 
and hence the event rate in the 
detector keeps increasing until this value of $\stch$ is reached. 
Beyond this limiting value of $\stch$, effect of earth 
matter in $P_{e\mu}$ falls and the event rate for 
the neutrinos, for normal hierarchy, and antineutrinos, for 
inverted hierarchy, decreases. However, note that this 
value of $\stch$ is already disfavored from the CHOOZ data
\cite{chooz,limits}. 
 
From the left-hand panel of 
Fig. \ref{fig:rates} we note that for $\gamma=500$ and 
$\stch=0.05$, the predicted number of neutrino
events for normal hierarchy 
is 101, while that for inverted hierarchy is only 3. This 
implies that if the normal hierarchy was true, we could 
comprehensively rule out the wrong inverted hierarchy. 
For smaller boost factors $\gamma$, 
the difference between the number of events decreases; 
however, we could still determine 
the neutrino mass hierarchy if $\stch$(true) is 
not very small. We will perform a statistical analysis 
of projected data in the following sections and 
present our results on the hierarchy sensitivity in section 7.
Instead of the neutrino, one could work with the $\anue$ 
\bb{} and achieve similar sensitivity to the neutrino mass 
hierarchy. Note that while the interaction cross section 
for the $\anue$s are much smaller,
the flux itself is larger owing to the (assumed) 
larger number of 
decays per year for $^8Li$. Thus the statistics 
expected in both the neutrino as well as the 
antineutrino channel is comparable and, as we will see, 
the hierarchy sensitivities for the $\nue$ and 
$\anue$ \bb are hence similar. 

As discussed above, resonant matter effects in the 
neutrino (antineutrino) channel for the normal (inverted) 
hierarchy, results in substantial enhancement in the 
observed number of events. In particular, we note from 
Fig. \ref{fig:rates} that the number of events in the 
near-resonant channels 
depend strongly on the value of $\stch$, since 
the extent of matter effects 
is dictated directly by $\theta_{13}$. We can see 
from the figure that the dependence of the event rate on 
$\stch$ is much enhanced due to matter effects. 
Therefore, if the true 
hierarchy was normal (inverted), we could use the 
neutrino (antineutrino) \bb in the proposed experiment to 
measure/constrain the mixing angle $\theta_{13}$. 

In Fig. \ref{fig:deg} we show the effect of $\delta_{CP}$
on the variation of the 
measured rate with the value of the mixing angle $\theta_{13}$, 
for four different baselines 732 km, 4000 km, 7152 km and 10000 km.
The left-hand panel shows the results for the $\nue$ $\beta$-beam
assuming a normal mass hierarchy, 
while the right-hand panel gives the 
same for the $\anue$ $\beta$-beam flux with inverted hierarchy.
For all cases in any panel, we have considered  a 
50 kton magnetized iron calorimeter as the far detector
and the same flux created at the 
source. 
Each  curve gives the sets of values of 
$\{\stch,\delta_{CP}\}$ which give the same observed rate 
in the detector as the set $\{\stch=0.04,\delta_{CP}=0^\circ\}$. 
In other words, if the true value of $\stch$ and 
$\delta_{CP}$ were 0.04 and $0^\circ$ respectively, then every  
point on a given curve would also be a solution for that 
experiment. It is clear from this figure that combining 
results from experiments at different baselines helps 
solve/reduce the problem. However, 
the most important issue exemplified here is the fact that 
for the baseline 7152 km, which is the CERN-INO distance, 
the effect of the unknown value of $\delta_{CP}$ on 
the measurement of $\theta_{13}$ and the mass hierarchy,
is very small. 
This happens because this  
distance corresponds to a near-magic baseline for which, as 
noted earlier, the $\delta_{CP}$ dependent terms are 
almost vanishing.

\subsection{Backgrounds}

The possible backgrounds in a $\nue$ \bb experiment\footnote{The 
discussion concerning the $\anue$ \bb is similar 
and hence is not repeated  
here.} using $\mu^-$
as an oscillation signal  
come from 
neutral current events such as
\be
\nu_x + d(u) &\rightarrow& \nu_x + d(u)
\\
\nu_x + d(u) &\rightarrow& \nu_x + d(u) + q\bar{q}
\ee
and $\nue$ charged current events
\be
\nue + d &\rightarrow& e^- + u \; {\rm or} \; c \;({\rm Cabibbo
\; suppressed})
\\
\nue + d &\rightarrow& e^- + u + q\bar{q}
\ee
The quarks  in the final state could produce mesons
as a part of the hadronic junk. These mesons may then decay 
producing secondary muons, giving rise to a  signal which 
might constitute a possible background. 

As discussed earlier, INO-ICAL will have 6 cm thick iron plates. 
Such a dense tracking detector will have  
excellent muon/pion and muon/electron separation 
capability in the energy range we are working with. 
The simulations carried out by the INO collaboration 
have shown that after the standard kinematical cuts are imposed,
the electrons do not give any signal at all, while in 99\% of the
cases, the pions and kaons get absorbed very quickly in the iron
for the energy range of interest to us and therefore do not hit
enough RPCs to give a signal.  At the energies of  the
$\beta$-beams  considered here, production cross section of $D$
mesons (also Cabibbo suppressed) is small and they do not
constitute a problem for the experiment.  The associated strange
or charm production is also highly suppressed at these energies.
In addition, the fact that the detector will have a charge
identification capability means that secondary $\mu^+$ produced
can be safely discarded, reducing the background even further.
Therefore, in this analysis we do not consider any backgrounds
coming from either the neutral current events or charged current
events of $\nue$\footnote{Mesons produced in neutral current 
processes are degraded in energy. Note that backgrounds from 
these mesons are very important in the 
case of the neutrino factory. However, since our relevant energy 
range is lower, the mesons produced 
in each event are much lower in 
energy and hence can be easily rejected by putting suitable cuts. In 
our analysis we have assumed cuts that are stringent enough
to completely reject these backgrounds.}.

Since the oscillation probability $\nue \rightarrow \nutau$ is
about the same as that for $\nue \rightarrow \numu$, we expect
almost as many $\nutau$ arriving at the detector as $\numu$. The
$\tau^-$ produced through charged current interaction may decay
producing secondary $\mu^-$ with a branching ratio of 17.36\%.
But, the $\tau$ threshold (3.5 GeV) is high and the production
cross section  suppressed compared to $\mu$. So we do not expect
any significant background from this source either.  We have
estimated the number of secondary muons produced from the
$\nutau$ component of the beam.  For $\stch=0.01$, we expect
0.008, 0.061 and 0.2 muon events per year respectively, for
$\gamma=250$, 350 and 500.  In addition, these secondary muons
will be severely degraded in energy and therefore can be
eliminated through energy cuts. We therefore neglect the
backgrounds  from this source as well.

\section{Details of the Statistical Method}

In order to quantify the sensitivity of this \bb
experimental set-up 
to the mixing angle $\theta_{13}$ and $sgn(\ma)$, we perform 
a statistical analysis of the ``data'' generated in INO-ICAL, 
assuming certain true values of the parameters. Since the number 
of events expected in the detector might be very small, depending on 
how small the true value of $\stch$ is, we define a $\chi^2$ 
assuming Poissonian distribution for the error as,
\be
\chi^2(\{\omega\}) = \min_{\xi_k}\left[2\left(\tilde N^{th}-N^{ex} 
-N^{ex} \ln \frac{\tilde N^{th}}{N^{ex}}\right) +
\sum_{k}\xi_k^2\right ]~.
\label{eq:chipull}
\ee
In Eq. (\ref{eq:chipull}), $\{\omega\}$ is a set of 
oscillation parameters,
$N^{ex}$ is the observed number of events, while 
the systematic errors
in the data and the theory are accounted for through the 
set of ``pulls''
$\{\xi_k\}$,
where $k$ runs over the different systematic uncertainties
involved.  The pulls are defined in such a way that the number
of expected events, $N^{th}$, corresponds to $\xi_k = 0$.
$\tilde{N}^{th}$ is the number of events when the systematic
errors are included in a manner such that the effect of the
$k$-th uncertainty at the $\pm1\sigma$ level is picked up when $\xi_k =
\pm 1$:
\be
\tilde{N}^{th}(\{\omega\},\{\xi_k\}) = N^{th}(\{\omega\}) \left[
1+ \sum_{k=1}^K \pi^k \xi_k \right] + {\cal O}(\xi^2_k)~,
\label{eq:rth}
\ee
In Eq. (\ref{eq:rth}) 
${\tilde N}^{th}(\{\omega\},\{\xi_k\})$ 
has been expanded in powers of
$\xi_k$, keeping only linear terms.  The quantities $\pi^k$ give the
fractional rate of change of $N^{th}(\{\omega\})$ due to the $k$th
systematic error. In our analysis, we include systematic 
uncertainties  in the normalization of the \bb flux 
at the source, the error in the cross section and the detector 
systematic uncertainty. For the 
flux normalization error we adopt a total uncertainty of 2\%.
The neutrino-nucleon cross sections, especially at 
large energies and for massive target nuclei, are 
known to have large uncertainties. 
We include a 10\% error coming from the uncertainty in the 
interaction cross section.  
A total detector systematic uncertainty of 2\% is also 
included. 
We do not include any 
systematic error related to the shape of the energy spectrum 
since we work only with the energy integrated total rates, for 
which such types of errors would be negligible.

Unless otherwise stated, the data are always generated at the 
benchmark values of the oscillation parameters $|\ma|$, $\ms$, 
$\sta$, $\sss$ and $\delta_{CP}$ given in Table \ref{tab:true}. 
The true 
value of $\theta_{13}$ as well as $sgn(\ma)$ 
will be allowed to change, since 
these are the parameters under study in this experimental set-up 
of interest. Throughout the paper we will use the notation 
$\stcht$ as the true value of this parameter chosen by 
Nature, and $\stch$ as the fitted value. 
The data, simulated with the (assumed) true
values of the oscillation parameters, 
are fitted   
by calculating $N^{th}(\{\omega\})$
for any set of values for the 
oscillation parameters. 
We next calculate the 
chi-square function $\chi^2(\{\omega\},\{\xi_k\})$
for every possible $\tilde{N}^{th}(\{\omega\},\{\xi_k\})$, 
obtained by varying the 
oscillation parameters and $\xi_k$. 
We first minimize this $\chi^2(\{\omega\},\{\xi_k\})$ function 
with respect to the pulls $\xi_k$.
This 
$\chi^2(\{\omega\})$ (given by Eq. (\ref{eq:chipull}))
is next minimized with respect 
to all the oscillation parameters\footnote{
In order to save computer time, 
we keep $\ms$ and $\sss$ fixed 
in our analysis. There is two-fold motivation for this 
approximation. Firstly, these parameters, especially $\ms$, 
is expected to be measured with a high precision in 
the proposed solar and reactor neutrino experiments 
\cite{solarprecision}. Secondly, since we are working 
at a near-magic baseline, we do not expect 
any significant impact of these parameters 
on our results.}, 
which are permitted to assume all 
allowed values in calculating $N^{th}(\{\omega\})$. 

It is expected that 
the parameters $|\ma|$ and $\sta$ would be fairly well determined 
by the long baseline experiments T2K \cite{t2k} and NO$\nu$A
\cite{nova}. In order to take into account this information, 
which should be available by the time the \bb facility comes up
at CERN, we add to our $\chi^2$ the ``prior'' function such that 
\be
\chi^2_{total} = \chi^2 + \chi^2_{prior}~,
\ee
\be
\chi^2_{prior} = \left (\frac{|\Delta m^2_{31}|-
|\mat|}{\sigma(\Delta m^2_{31})} \right )^2 + 
\left (\frac{\sta-\stat}{\sigma(\sta)} \right )^2~,
\ee
where $\mat$ and $\sta$(true) are taken as the benchmark values in 
Table \ref{tab:true}.
We will assume that the $1\sigma$ error on these parameters 
would be reduced to $\sigma(\Delta m^2_{31})=1.5\%$ and 
$\sigma(\sta)=1.0\%$ \cite{huber10}. We minimize the 
total function $\chi^2_{total}$ with respect to the 
oscillation parameters and present the C.L. results 
assuming as a convention that $n\sigma$ corresponds 
to 
\be
\Delta \chi^2_{total} = n^2~,
\ee
where $\Delta \chi^2_{total}$ is the 
minimum $\chi^2_{total}$ obtained in the analysis
for a given value of the 
parameter $\stcht$ and/or the mass hierarchy.

\begin{figure}[t]
\begin{center}
\includegraphics[width=9.0cm, height=8.0cm]{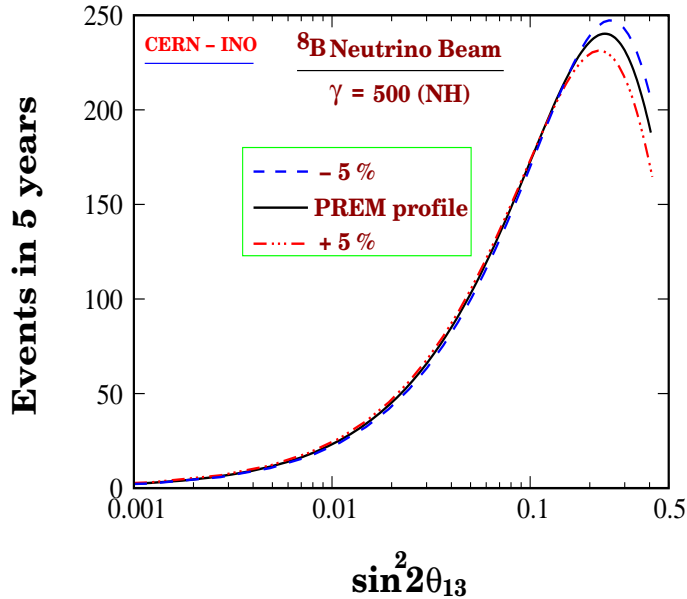}
\caption{\label{fig:premerror}
Number of events versus $\stch$ for the $\nue$ \bb with 
$\gamma=500$, showing the impact of uncertainty 
in the earth matter density profile. The solid curve corresponds 
to the PREM matter density profile, the dashed (dot-dashed) curve is for 
the PREM density scaled  by -5\% (+5\%).
}
\end{center}
\end{figure}

Note that we have not taken into account any error coming from 
the uncertainty in the matter density. Fig. \ref{fig:premerror}
shows our justification for this. This figure shows 
the expected event rate for the $\nue$ \bb as a function 
of $\stch$. The solid curve corresponds 
to the PREM profile, while the dashed and 
dot-dashed curves are for PREM densities scaled by -5\% and 
+5\% respectively.  We can see that the effect of the 5\% 
uncertainty in the matter density profile shows up in the 
event rate only for values of $\stch$ already disfavored by 
the CHOOZ experiment. In particular, we note that 
for $\stch \ltap 0.1$ the effect of the matter density 
uncertainty is negligible.

\section{Determining $\sin^22\theta_{13}$}

We begin by presenting results on how well $\stch$ can be
determined by the set-up we are proposing. We divide
our results into two subcategories. First we assume that
$\stch$(true)$=0$ and expound the possibility of improving the
upper bound on this parameter. This is what we would call the
sensitivity of the experiment to $\stch$. Next we will assume
that $\stch$(true) is large enough to give a positive signal in
the experiment. In that case, we will probe quantitatively how
precisely this parameter could then be determined.
All results are presented after marginalizing over
$|\ma|$, $\sta$, and $\delta_{CP}$, as described in section 5.

\subsection{Sensitivity to  $\sin^22\theta_{13}$}

\begin{figure}
\includegraphics[width=8.5cm, height=8.0cm]{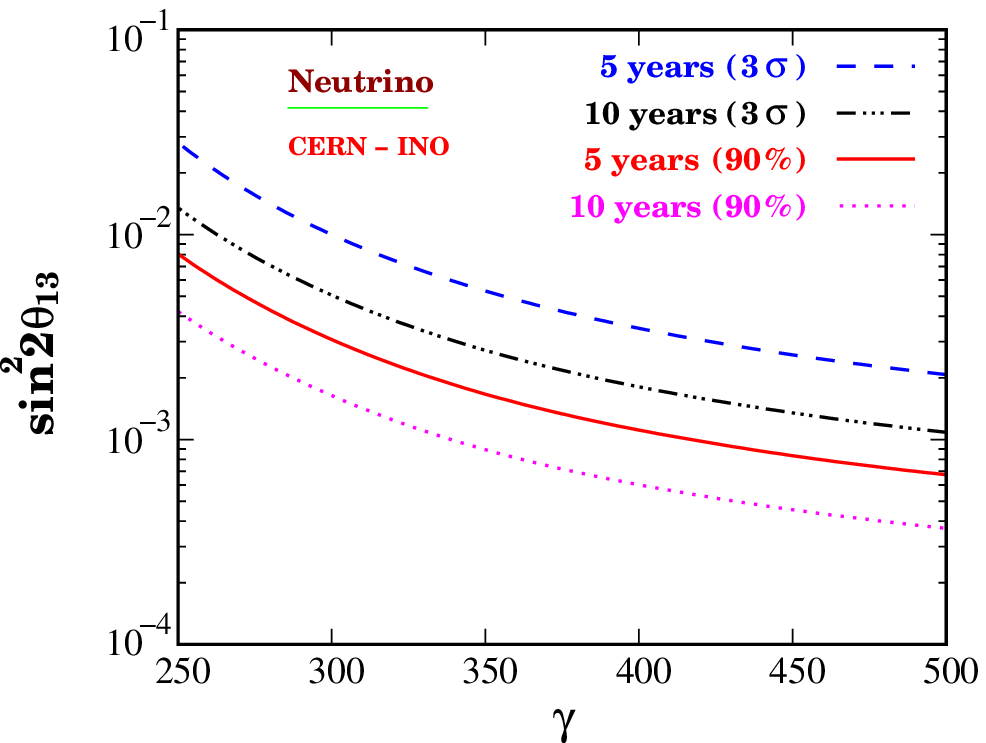}
\vglue -8.0cm \hglue 8.5cm
\includegraphics[width=8.5cm, height=8.0cm]{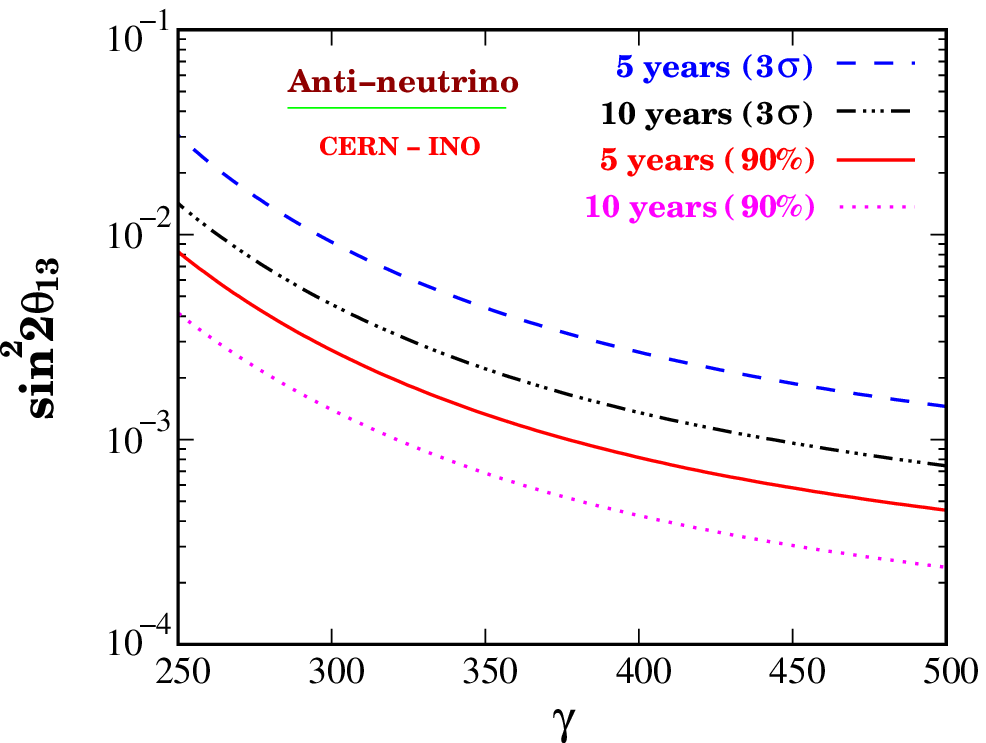}
\caption{\label{fig:sensth13}
The upper limit on $\stch$ that can be imposed by the CERN-INO \bb 
set-up with 5 years and 10 years of running. The left-hand
panel shows the $\stch$
sensitivity expected for the neutrino mode assuming 
normal hierarchy to be true, while the right-hand panel 
shows the sensitivity for the antineutrino mode and an
inverted hierarchy. 
}
\end{figure}


We define the sensitivity to $\theta_{13}$ as 
the minimum value of $\stch$ which this 
experiment would be able to distinguish from $\stcht=0$. 
The results of our analysis are presented in Fig. \ref{fig:sensth13}, 
where we show the upper bounds on $\stch$ that
this experiment can impose
at the 90\% C.L. (solid curve for 5 years running and 
dotted curve for 10 years 
running) 
and \sig{} C.L. (long dashed curve for 5 years running and 
dashed-dotted curve for 10 years 
running). The left-hand panel is for a $\nue$ \bb assuming 
that normal hierarchy is the true hierarchy, while the 
right-hand panel is for an $\anue$ \bb with the assumption that
the inverted hierarchy is true. 
We show 
our results as a function of the Lorentz boost $\gamma$. 
The 90\% and $3\sigma$ C.L. upper bounds that we can  
impose on $\stch$ are displayed in Table \ref{tab:sens_th13} 
for the three benchmark values of $\gamma$ and with five and ten 
years of running of the experiment in the neutrino (columns 3 and 4)
and in the antineutrino (columns 5 and 6) mode. 
In the Table we show the $\stch$ sensitivity 
limits for the conservative detector efficiency of 60\% for
$\mu^\pm$ as well 
as for 80\% efficiency. 
%
\begin{table}[t]
\begin{center}
\begin{tabular}{|c|c|c|c|c|c|c|} \hline
&& & \multicolumn{2}{|c|}{{\rule[0mm]{0mm}{6mm}Neutrino Beam (NH true)}} 
& \multicolumn{2}{|c|}{\rule[-3mm]{0mm}{6mm}{Anti-neutrino Beam (IH true)}}
\cr \cline{4-7} 
Detection & Years & $\gamma$ & {\rule[0mm]{0mm}{6mm}90\% C.L.} 
& \sig{} C.L. & 90\% C.L. & \sig{} C.L. \cr
Efficiency& & & & & &  \cr 
\hline
\hline
\multirow{6}{*}{60\%}&   
& 250 & $~8.1\times 10^{-3}~$ & $2.8\times 10^{-2}$ 
& $~8.3\times 10^{-3}~$ & $3.1\times 10^{-2}$ \cr
& 5 years & 350 & $1.6\times 10^{-3}$ & $5.2\times 10^{-3}$ 
& $1.3\times 10^{-3}$ & $4.3\times 10^{-3}$ \cr
&        & 500 & $6.7\times 10^{-4}$ & $2.1\times 10^{-3}$ 
& $4.5\times 10^{-4}$ & $1.5\times 10^{-3}$ \cr
\cline{2-7}
&        & 250 & $4.2\times 10^{-3}$ & $1.4\times 10^{-2}$ 
& $4.1\times 10^{-3}$ & $1.4\times 10^{-2}$ \cr
& 10 years & 350 & $8.7\times 10^{-4}$ & $2.7\times 10^{-3}$  
& $6.7\times 10^{-4}$ & $2.2\times 10^{-3}$  \cr
&        & 500 & $3.7\times 10^{-4}$ & $1.1\times 10^{-3}$ 
& $2.4\times 10^{-4}$ & $7.4\times 10^{-4}$  \cr
\hline
\hline
\multirow{6}{*}{80\%}&        
& 250 & $~6.3\times 10^{-3}~$ & $2.1\times 10^{-2}$
& $~6.2\times 10^{-3}~$ & $2.2\times 10^{-2}$  \cr
&5 years & 350 & $1.3\times 10^{-3}$ & $4.0\times 10^{-3}$
& $9.8\times 10^{-4}$ & $3.2\times 10^{-3}$ \cr
&        & 500 & $5.3\times 10^{-4}$ & $1.6\times 10^{-3}$
& $3.5\times 10^{-4}$ & $1.1\times 10^{-3}$ \cr
\cline{2-7}
&        & 250 & $3.4\times 10^{-3}$ & $1.0\times 10^{-2}$
& $3.1\times 10^{-3}$ & $1.1\times 10^{-2}$ \cr
&10 years & 350 & $6.9\times 10^{-4}$ & $2.1\times 10^{-3}$
& $5.1\times 10^{-4}$ & $1.6\times 10^{-3}$  \cr
&        & 500 & $2.9\times 10^{-4}$ & $8.4\times 10^{-4}$
& $1.8\times 10^{-4}$ & $5.7\times 10^{-4}$  \cr
\hline
\end{tabular}
\caption{\label{tab:sens_th13}
Upper limit on $\stch$ which  can be imposed 
with 5 years and 10 years data respectively, at the 90\% and \sig{} C.L., 
for different benchmark values of $\gamma$.
Columns 3 and 4 show the sensitivity using the $\nue$ beam 
and assuming normal hierarchy as true and 
columns 5 and 6 are the sensitivity possible with the 
$\anue$ beam and assuming inverted hierarchy as true.
Results are shown for 60\% and 80\% $\mu^\pm$ detection efficiency.
}
\end{center}
\end{table}
%
This can be compared with the sensitivity we expect from the 
other forthcoming/planned experiments. 
At the 90\% C.L., the T2K and NO$\nu$A experiments 
are expected to constrain $\stch < 2.3\times 10^{-2}$ and 
$\stch < 2.4\times 10^{-2}$ 
respectively \cite{huber10}. 
The Double-Chooz experiment is likely to
push the limit down 
to $\stch < 3.2\times 10^{-2}$ at 90\% C.L., which could be 
improved to $\stch < 9\times 10^{-3}$ 
by the Reactor-II set-up \cite{huber10}.
The ``discovery potential''\footnote{The authors 
define the ``discovery potential'' as the minimum value of 
$\stch$(true) for which $\stch=0$ gives a 
$\Delta\chi^2=9$.} of the 
CERN-MEMPHYS project for five years of running 
of both the \bb and the SPL super-beam  
set-ups considered alone is 
$\stch \ltap 5\times 10^{-3}$ at \sig{} \cite{cernmemphys}.
With a combined data set in MEMPHYS 
for five years running in \bb and five years in 
the SPL super-beam, the discovery limit could be improved to  
$\stch \ltap 3.5\times 10^{-3}$ at \sig{} \cite{cernmemphys}.
The best limit, obtained 
after a thorough optimization of the \bb experimental options 
for the $\theta_{13}$ sensitivity reach   
in \cite{betaoptim} (see also \cite{betaoptim2}),
is $\stch \ltap 1.5\times 10^{-3}$ at \sig,
for what the authors label as set-up 1\footnote{This corresponds to
successive 8 year runs of a 
\bb in the neutrino and antineutrino mode (total
16 years) 
with $\gamma=200$, $L=520$ km 
and a water \u{C}erenkov detector with 500 kton 
fiducial mass.} and set-up 2\footnote{This corresponds to
successive 8 year runs of a 
\bb in the neutrino and antineutrino mode (total
16 years) 
with $\gamma=500$, $L=650$ km  
and 50 kton of totally 
active scintillator detector.}. Therefore, 
for the experimental set-up 
with the \bb source in CERN and INO-ICAL as the far detector 
the projected $\stch$ sensitivity is comparable to, if not 
better, than most of the other planned schemes.
In fact, the sensitivity reach of this experiment is 
even better than that expected for an entry level 
Neutrino Factory (NuFact-I) \cite{sbvsnufact}. 
Only a NuFact-II with $L=7500$ km \cite{magic} corresponding to the 
magic baseline, could give a $\stch$ sensitivity significantly 
better than what one can achieve with the CERN-INO \bb set-up 
\cite{magic,optimnufact}.

We reiterate that the tremendous $\stch$ 
sensitivity of the CERN-INO 
\bb set-up comes not really from high \bb flux, but 
from the fact that one has near-resonant matter 
effects at the near-magic baseline. While the latter feature
smothers the problems arising from clone solutions, the former 
lends the experiment an extra sensitivity due to the 
dependence of matter effects on $\stch$. This is why 
the $\stch$ sensitivity depends on the true hierarchy. We have 
checked that if we run the experiment for 10 years in the 
neutrino mode alone and 
with inverted hierarchy assumed as true, then the 
sensitivity we expect is a meagre
$\stch > 1.8\times 10^{-2}$ at the 90\% 
C.L. Therefore, a hierarchy independent $\stch$ sensitivity
can in principle be obtained after runs in both the 
neutrino and the antineutrino mode, unless one is very lucky.
We will return to this issue again in section 8.



\subsection{Precision in measurement of $\sin^22\theta_{13}$}



\begin{figure}[p]
\begin{center}
\includegraphics[width=16.0cm,height=9.0cm]{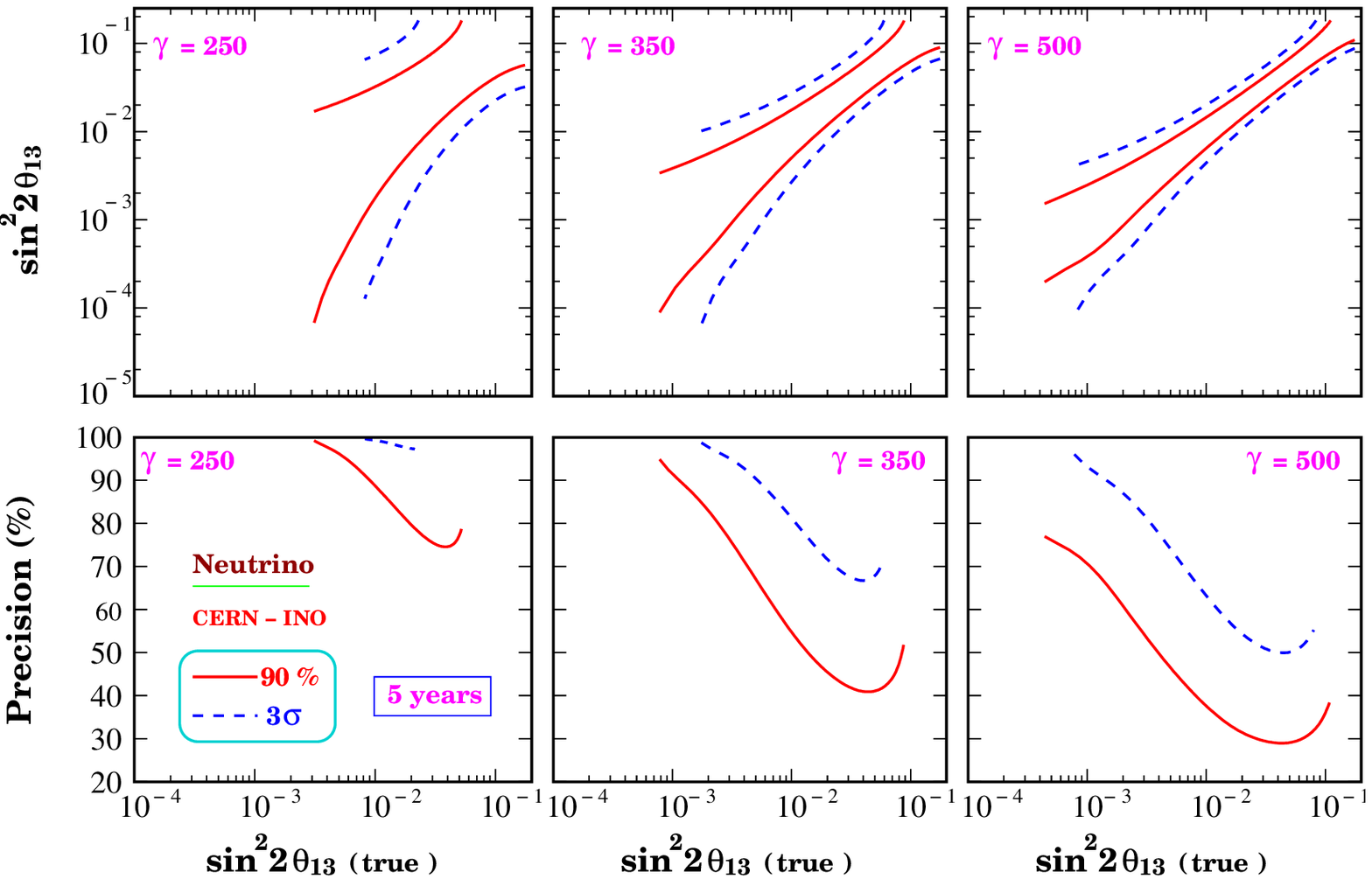}
\caption{\label{fig:precision}
Precision in the measurement of $\stch$ expected in 
the CERN-INO \bb experiment with 5 years running in 
the neutrino channel assuming normal hierarchy. 
Upper panels show the 
90\% and \sig{} band of ``measured"
values of $\stch$ 
for every $\stch$(true). 
Lower panels show the corresponding 
90\% and \sig{} C.L. precision as
defined in the text, as a function of $\stch$(true).  
}
\end{center}
\begin{center}
\includegraphics[width=16.0cm,height=9.0cm]{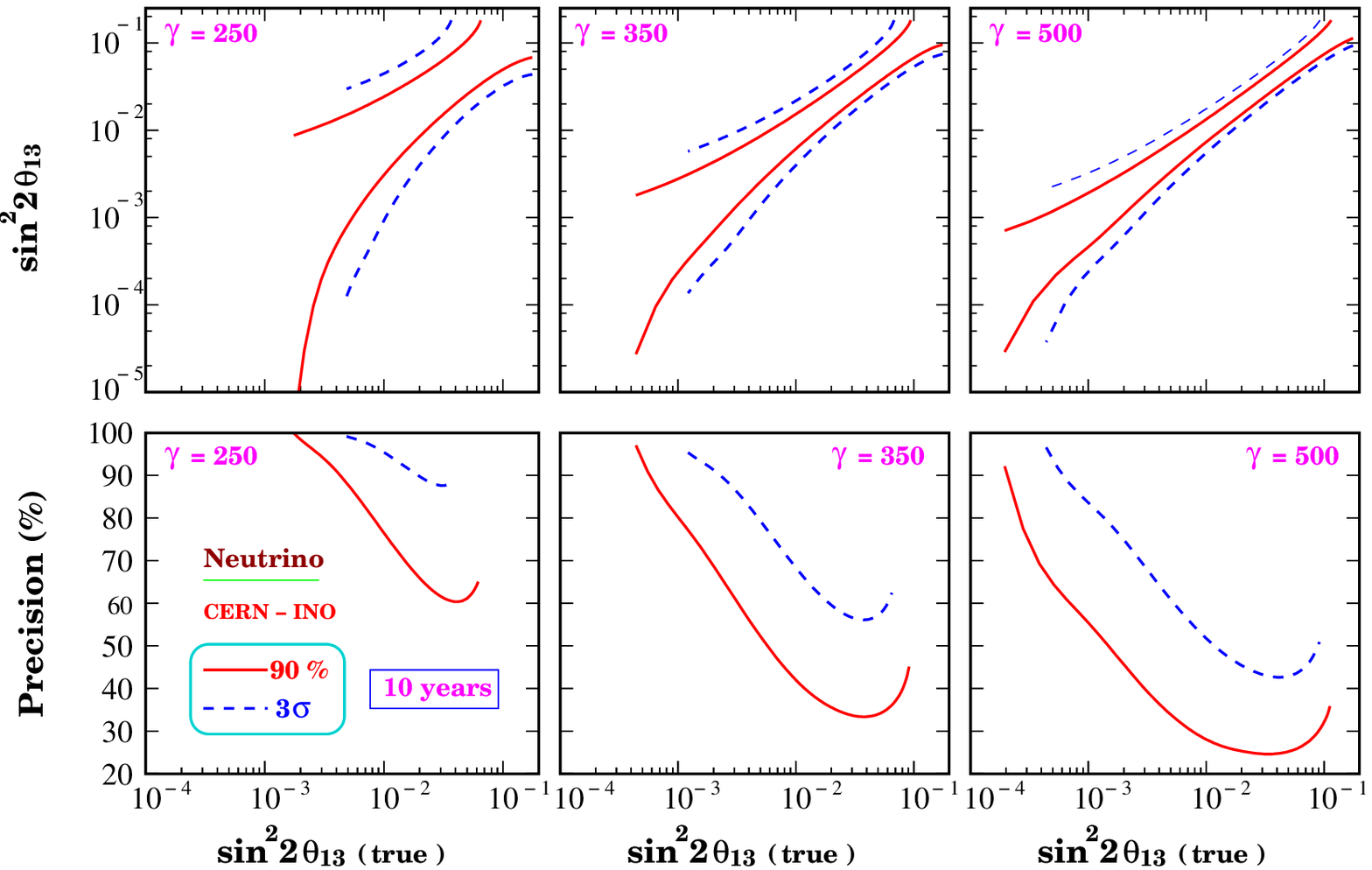}
\caption{\label{fig:precision10}
Same as Fig. \ref{fig:precision} but for 10 years of 
running.
}
\end{center}
\end{figure}

In this subsection we assume that $\stch$(true) is 
large enough such that, for a given boost $\gamma$,
we have a positive signal in the ICAL detector. For each 
value of $\stch$(true), varied between the limiting value 
obtained in the previous subsection and the current upper limit, 
we generate the data and 
fit it back using the $\chi^2$ analysis to give us 
the range of ``measured" values of $\stch$. We plot in 
the upper panels of Figs. \ref{fig:precision} 
and \ref{fig:precision10}, the 
90\% and \sig{} C.L. bands of $\stch$ as a function 
of $\stch$(true) for 5 years and 10 years run respectively. 
The left, middle and right-hand panels are 
for $\gamma=250$, 350 and 500, respectively. Plots are shown for
the $\nue$ \bb assuming normal hierarchy as true.  In order to
quantify the precision which can be obtained on $\stch$, we
define, for any value of $\stch$(true),
\be
{\rm Precision} = \frac{{\stch}({max}) - {\stch}({min})}
{{\stch}({max}) + {\stch}({min})} \times 100\,\%~,
\ee
where ${\stch}({max})$ and ${\stch}({min})$ are 
respectively the largest and 
smallest values of $\stch$ allowed at the given C.L. We 
show in the lower panels of Figs. \ref{fig:precision}
and \ref{fig:precision10},
the variation of this precision function with 
$\stch$(true) for the 90\% and \sig{} C.L. for 5 years
and 10 years of 
running respectively, in the
neutrino channel assuming normal hierarchy. 
Similar results are expected if we run the experiment in the 
antineutrino mode for inverted hierarchy.

As expected, the precision in the $\stch$ measurement 
improves as $\stcht$ increases. However,
a very interesting feature in the plots of Fig. \ref{fig:precision}
is the emergence of minima in the precision, thereby 
implying that for values of $\stch$(true) beyond this 
limiting value, the precision actually deteriorates.

\section{Sensitivity to $Sgn(\ma)$}

\begin{figure}
\includegraphics[width=8.5cm, height=8.0cm]{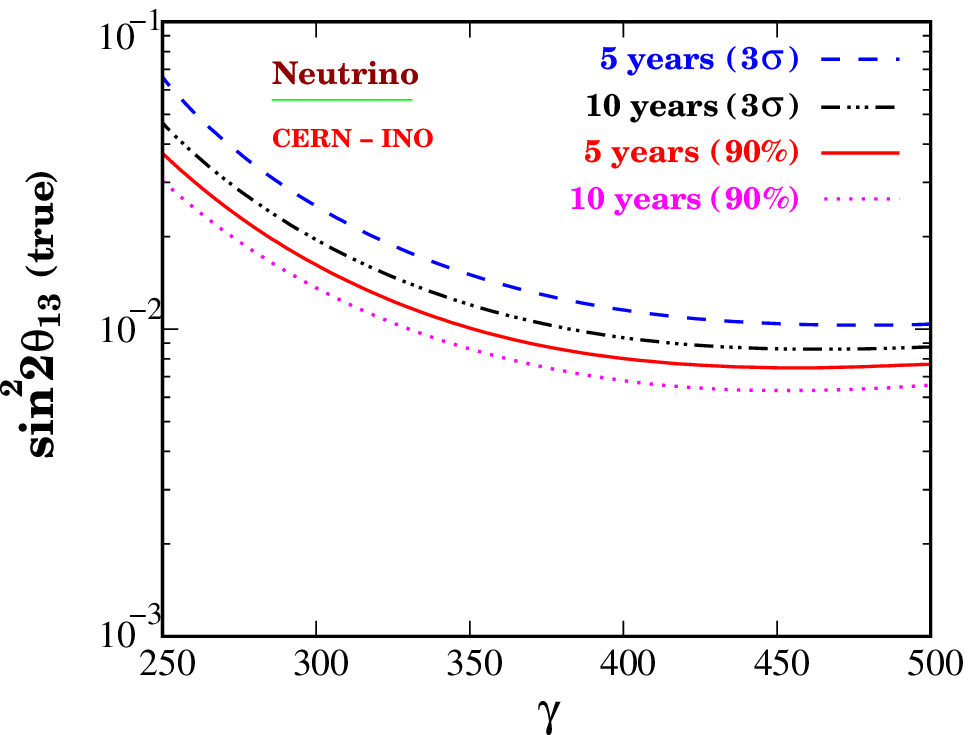}
\vglue -8.0cm \hglue 8.5cm
\includegraphics[width=8.5cm, height=8.0cm]{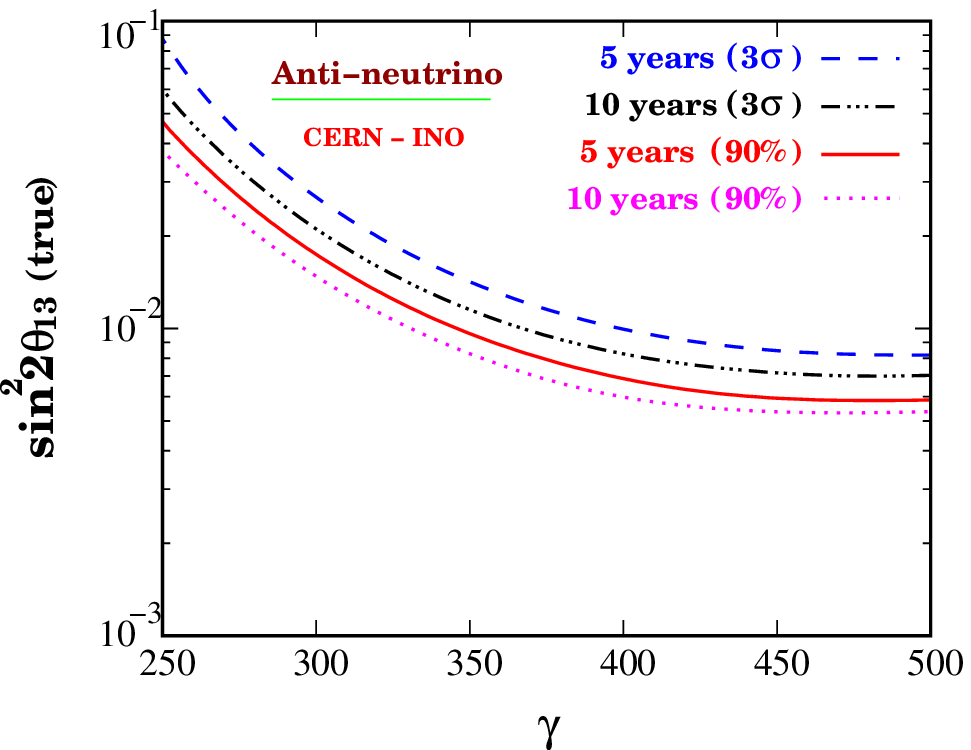}
\caption{\label{fig:hier}
The minimum value of $\stch$(true) as a function of the boost factor $\gamma$
at which the wrong hierarchy can be disfavored at the 
90\% 
and \sig{} C.L. Results are presented for 5 years and 10 years of
running in the $\nue$ (left-hand panel) and $\anue$ (right-hand
panel) mode, if the true hierarchy was normal and inverted
respectively.  } 
\end{figure}

We next turn our attention to the 
determination of the neutrino mass hierarchy. 
If the true value of $\theta_{13}$ has indeed  
been chosen to be large by Nature, we can expect 
sizable matter effects in this experimental set-up, 
giving us a handle on the mass hierarchy.
In our analysis, we 
assume a certain true mass hierarchy and generate the 
prospective data in INO-ICAL for the benchmark oscillation 
parameter values given in Table \ref{tab:true} 
and for different values of $\stcht$. 
We next fit this data with the wrong hierarchy and calculate a
$\chi^2$, after marginalizing over $|\ma|$, $\sta$, $\delta_{CP}$
as well as $\stch$, as described in section 5.  The results of
this analysis are presented in Fig. \ref{fig:hier}.  The figure
shows, as a function of $\gamma$, the minimum value of
$\stch$(true) for which the wrong hierarchy can be disfavored at
the 90\% C.L. (solid curve for 5 years running and dotted curve for
10 years running) and \sig{} C.L.  (long dashed curve for 5 years
running and dashed-dotted curve for 10 years running). The
left-hand panel is for a $\nue$ \bb and assuming that normal
hierarchy is the true hierarchy, while the right-hand panel is
for an $\anue$ \bb with the assumption that the inverted
hierarchy is true. Of course, since large matter effects appear
in only either the neutrino or the antineutrino channel for a
given hierarchy, the sensitivity of the experiment to $sgn(\ma)$
would be zero if we used a $\anue$ beam when the true hierarchy
was normal or an $\nue$ beam when the true hierarchy was
inverted.

\begin{table}[t]
\begin{center}
\begin{tabular}{|c|c|c|c|c|c|c|} \hline
&& & \multicolumn{2}{|c|}{{\rule[0mm]{0mm}{6mm}Neutrino Beam (NH true)}} 
& \multicolumn{2}{|c|}{\rule[-3mm]{0mm}{6mm}{Anti-neutrino Beam (IH true)}}
\cr \cline{4-7}
Detection & Years & $\gamma$ & {\rule[0mm]{0mm}{6mm}90\% C.L.} 
& \sig{} C.L. & 90\% C.L. & \sig{} C.L. \cr
Efficiency& & & & & & \cr
\hline\hline
\multirow{6}{*}{60\%}& 
        & 250 & $~3.7\times 10^{-2}~$ & $6.6\times 10^{-2}$ 
& $~4.7\times 10^{-2}~$ & $8.7\times 10^{-2}$  \cr
&5 years & 350 & $9.7\times 10^{-3}$ & $1.5\times 10^{-2}$ 
& $9.2\times 10^{-3}$ & $1.4\times 10^{-2}$ \cr
&        & 500 & $7.7\times 10^{-3}$ & $1.0\times 10^{-2}$ 
& $5.9\times 10^{-3}$ & $8.2\times 10^{-3}$ \cr
\cline{2-7}
&        & 250 & $3.0\times 10^{-2}$ & $4.7\times 10^{-2}$ 
& $3.8\times 10^{-2}$ & $6.0\times 10^{-2}$ \cr
&10 years & 350 & $8.3\times 10^{-3}$ & $1.2\times 10^{-2}$  
& $8.1\times 10^{-3}$ & $1.1\times 10^{-2}$  \cr
&        & 500 & $6.6\times 10^{-3}$ & $8.7\times 10^{-3}$ 
& $5.4\times 10^{-3}$ & $7.0\times 10^{-3}$  \cr
\hline\hline
\multirow{6}{*}{80\%}&   
& 250 & $~3.4\times 10^{-2}~$ & $5.7\times 10^{-2}$
& $~4.3\times 10^{-2}~$ & $7.3\times 10^{-2}$  \cr
&5 years & 350 & $9.3\times 10^{-3}$ & $1.3\times 10^{-2}$
& $8.9\times 10^{-3}$ & $1.3\times 10^{-2}$ \cr
&        & 500 & $7.4\times 10^{-3}$ & $9.8\times 10^{-3}$
& $5.9\times 10^{-3}$ & $8.2\times 10^{-3}$ \cr
\cline{2-7}
&        & 250 & $2.8\times 10^{-2}$ & $4.2\times 10^{-2}$
& $3.5\times 10^{-2}$ & $5.4\times 10^{-2}$ \cr
&10 years & 350 & $8.1\times 10^{-3}$ & $1.1\times 10^{-2}$
& $7.8\times 10^{-3}$ & $1.0\times 10^{-2}$  \cr
&        & 500 & $6.6\times 10^{-3}$ & $8.5\times 10^{-3}$
& $5.4\times 10^{-3}$ & $6.9\times 10^{-3}$  \cr
\hline
\end{tabular}
\caption{\label{tab:hier_th13}
Minimum value of $\stch$(true) for which a CERN-INO beta-beam
experiment 
can rule out  the wrong hierarchy at the 90\% and \sig{} C.L. 
after 5 years and 10 years of running, 
for different benchmark values of $\gamma$.
Columns 3 and 4 show the sensitivity using the $\nue$ beam 
assuming normal hierarchy as true and 
columns 5 and 6 are for 
the $\anue$ beam assuming the inverted hierarchy as true.
Results are shown for 60\% and 80\% $\mu^\pm$ detection efficiency.
}
\end{center}
\end{table}

In Table \ref{tab:hier_th13} we show the limiting values of
$\stch$(true), for which the wrong hierarchy can be disfavored 
at the 90\% and \sig{} C.L. for three specific values of $\gamma$. 
We show the results for 5 years and 10 years of running of the 
experiment in the neutrino and the antineutrino mode. 
Table \ref{tab:hier_th13} shows the results for both 60\% and
80\% $\mu^\pm$ detection efficiency.  
This can be directly compared with the 
expected potential of the other 
planned/proposed long baseline experiments 
in discriminating between the 
right and the wrong hierarchy. The T2K experiment \cite{t2k}
has a baseline 
of only 295 km which entails almost negligible matter effects. The 
NO$\nu$A experiment \cite{nova}
on its own might find it hard to say anything conclusive
about the neutrino mass hierarchy, unless Nature conspires to 
arrange oscillation parameters in a conducive way. The 
so-called T2KK experimental 
set-up with the 4 MW beam from the JPARC facility and two 
detectors, one in Japan and another in Korea 
\cite{t2kk-I}\footnote{For an alternate 
T2KK proposal, see \cite{t2kk-II}.},
is expected to rule out the wrong hierarchy at the \sig{} 
C.L. if $\stcht > 5.5\times 10^{-2}$ for a total of 8 
years of running, 4 years in the $\nue$ and 4 
years in the $\anue$ mode. 
The 
CERN-MEMPHYS \bb and SPL superbeam 
proposals \cite{cernmemphys} 
again concern a baseline of only 130 km, 
which is too short to enable these beam experiments 
to decipher the mass hierarchy on their own. 
A higher $\gamma$ \bb 
from CERN to (say) Gran Sasso, with a baseline of 
732 km could have sensitivity to the mass hierarchy 
\cite{doninialter,betaoptim2}. 
It was 
shown in \cite{betaoptim} that for $\gamma=200$ and 
$L=520$ km, the wrong hierarchy could be disfavored at \sig{} C.L.
if $\stcht > 2\times 10^{-2}$ after 8 years of 
simultaneous running in the neutrino and antineutrino mode and 
with a 500 kton water detector. 
Therefore we see that
our sensitivity to the neutrino mass hierarchy can be 
well 
surpassed only by a Neutrino Factory, where the 
right normal hierarchy could be determined at \sig{} C.L.
for $\stcht > 3\times 10^{-4}$, if the experiment is done 
at the magic baseline $L=7500$ km.  

\begin{figure}
\includegraphics[width=8.5cm, height=8.0cm]{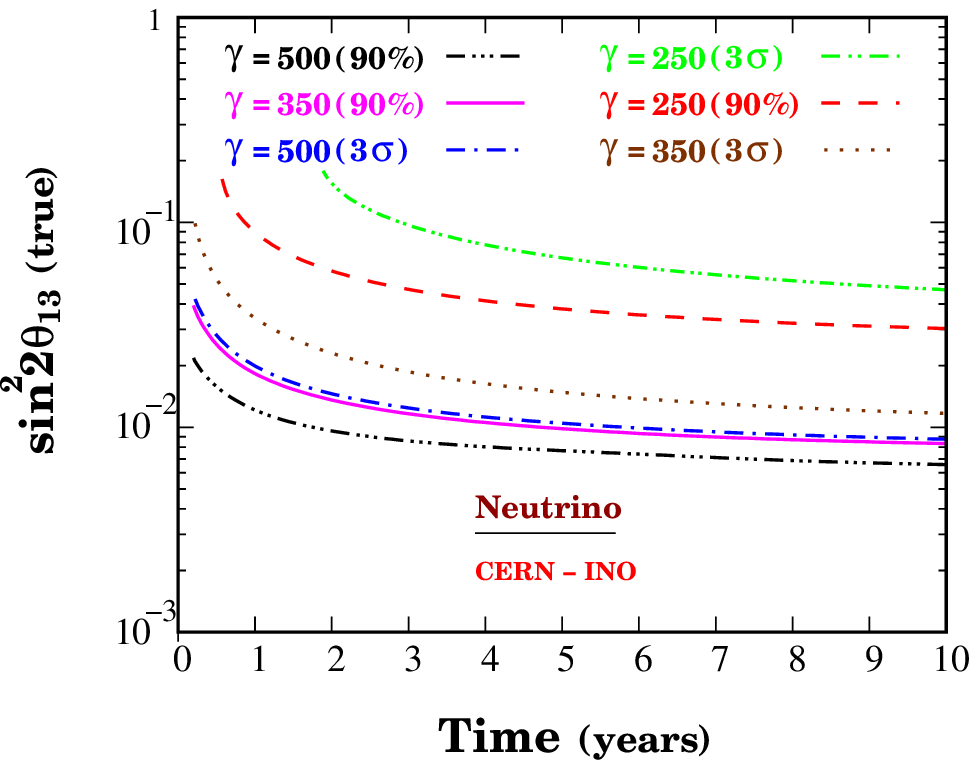}
\vglue -8.0cm \hglue 8.5cm
\includegraphics[width=8.5cm, height=8.0cm]{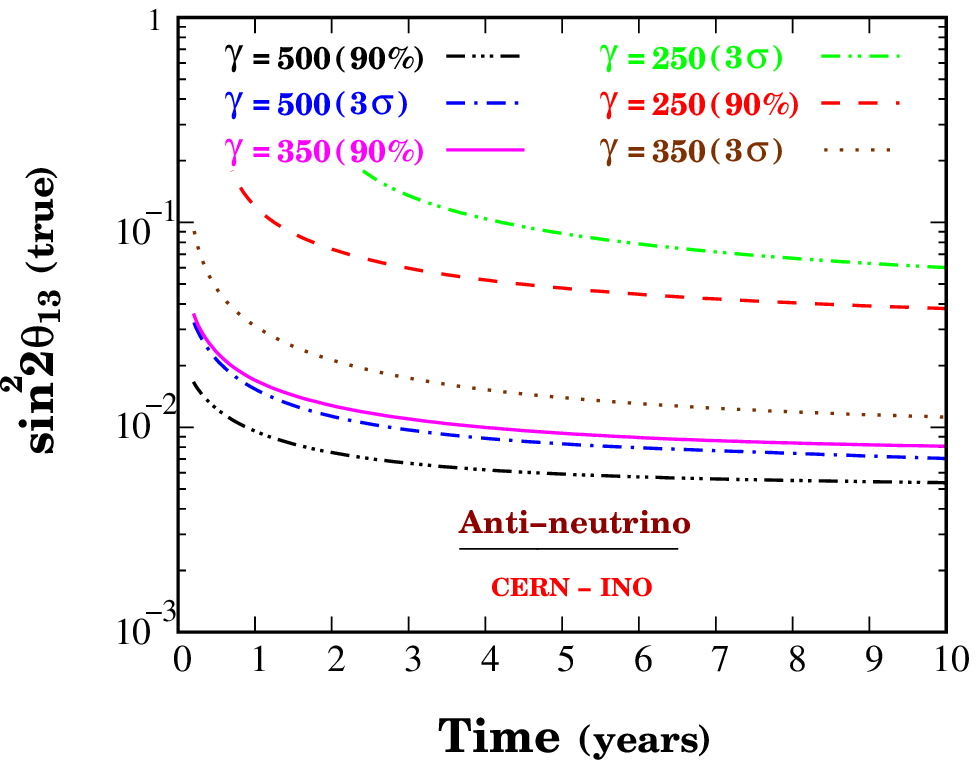}
\caption{\label{fig:hiertime}
The minimum value of $\stcht$ as a function of the running time necessary
to disfavor the wrong hierarchy at the 90\% and \sig{}
C.L. in the neutrino mode assuming normal hierarchy as true (left-hand 
panel) and in the antineutrino mode assuming inverted hierarchy
as true 
(right-hand panel).  Results 
are shown for three values of $\gamma$.
}
\end{figure}

\begin{figure}[t]
\includegraphics[width=8.5cm, height=8.0cm]{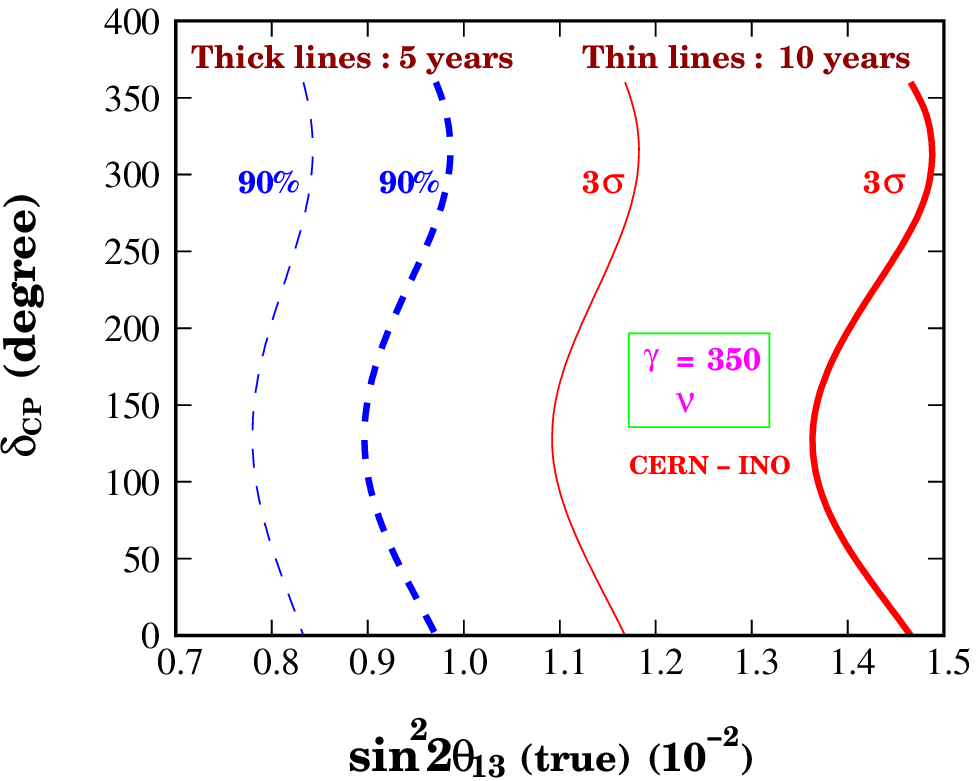}
\vglue -8.0cm \hglue 8.5cm
\includegraphics[width=8.5cm, height=8.0cm]{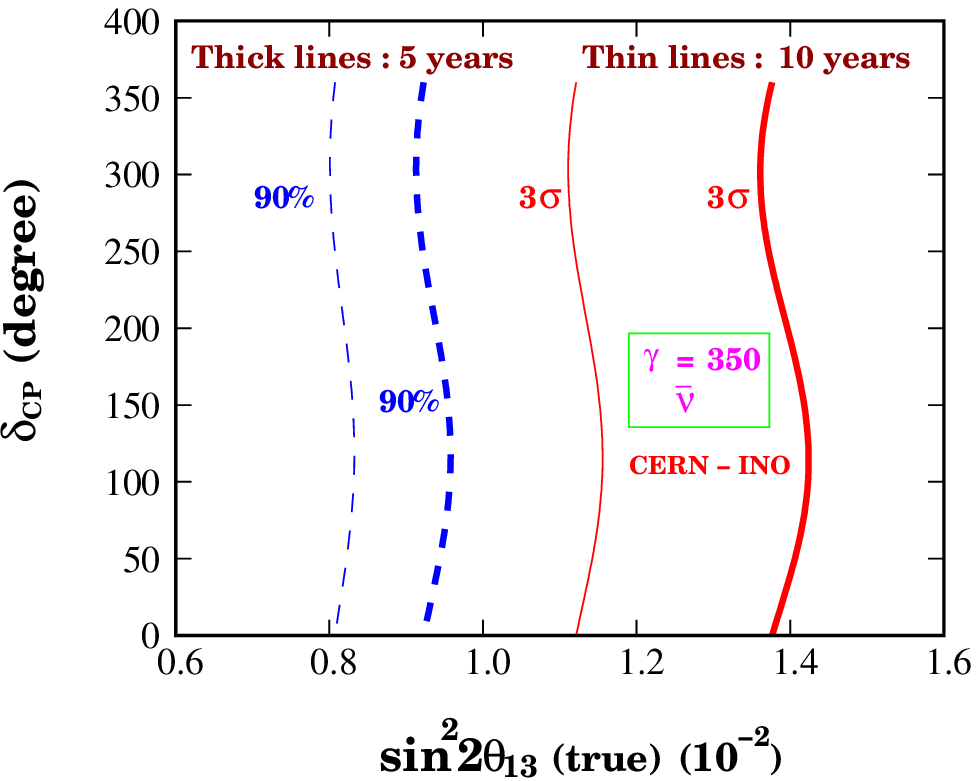}
\caption{\label{fig:hiercp}
The effect of $\delta_{CP}$ uncertainty 
on the limiting value of $\stcht$ for which the 
hierarchy determination will be feasible for $\gamma=350$. For
all  values to the 
right of the solid curves (dashed curves) it would be possible to 
rule out the wrong hierarchy at the \sig{} (90\%) C.L.
Thick (thin) lines are for 5 (10) years of running. Left-hand 
panel is for $\nue$ \bb assuming normal hierarchy is true,
while the right-hand panel is for $\anue$ \bb assuming 
true inverted hierarchy.}
\end{figure}

In Fig. \ref{fig:hiertime} we show the minimum $\stcht$ as
a function of the running time 
T (in years) needed
in this experimental set-up to discriminate between the wrong 
and the right hierarchy at the 90\% and \sig{} C.L. 
The left-hand panel is for a run in the $\nue$ mode 
with true normal hierarchy, while the right-hand panel shows the 
results for running in the $\anue$ mode with  
true inverted hierarchy. We show the results for the 
three chosen values of $\gamma$. 
We see, for example, 
that if $\stcht=5\times 10^{-2}$ and the normal hierarchy 
was true, then the wrong 
inverted hierarchy could be disfavored at the \sig{}
C.L. after just 9 (26) months of running in the $\nue$ 
mode with $\gamma=500$ (350).  

Fig. \ref{fig:hiercp} shows the impact of $\delta_{CP}$ 
on the hierarchy sensitivity of this experiment. The left-hand panel 
is for the neutrino channel with normal hierarchy as true, 
while the right-hand panel is for the antineutrino channel with 
inverted hierarchy as true. We show the 90\% (dashed curves) 
and \sig{} (solid curves) C.L. curves for 5 years (thick lines)
and 10 years (thin lines) running of the \bb with $\gamma=350$. 
We note that loss in hierarchy sensitivity due to the 
uncertainty in $\delta_{CP}$ is very marginal, owing to the 
fact that we are proposing this set-up at a near-magic baseline.
The best sensitivity to hierarchy comes for $\delta_{CP} \simeq 125$ 
in the neutrino mode and $\delta_{CP} \simeq 300$ in the antineutrino 
mode.

\section{Running the $\nue$ and $\anue$ $\beta$-Beam Simultaneously }

An exquisite feature of magnetized iron detectors is their charge
identification capability. Thus, if both the $\nue$ and the 
$\anue$ \bb were produced simultaneously at CERN, the oscillated 
$\numu$ could be measured distinctly from the oscillated 
$\anumu$ at INO\footnote{The $\nue$ and $\anue$ \bb fluxes 
could be produced in distinct bunches from the storage ring. 
Therefore, in principle, it should be straight-forward to 
distinguish the neutrino from the antineutrino events at 
any detector using the nano-second time resolution that most 
detectors possess. However, since ICAL will be magnetized, 
it could directly distinguish between the two channels irrespective
of the timing information.}. 
Since near-resonant matter effects appear only 
in either the neutrino or the antineutrino channel, this gives 
the experiment an added edge to unravel both the mass hierarchy
as well as $\theta_{13}$ simultaneously, within a much shorter time. 
The argument goes as follows.

Throughout the paper so far, we had been tacitly working in 
a scenario where we have just one type of \bb from CERN and 
the hierarchy was conducive to having large matter effect, the
key behind the excellent 
$\theta_{13}$ as well as 
mass hierarchy sensitivity. However, we have to bear in mind that 
in the absence of any prior knowledge of the true value of $\theta_{13}$ 
and/or the mass hierarchy, one would not know which of the two
modes -- 
neutrino or the antineutrino --
would be preferable for extracting maximal matter effects.
With luck, it might so happen 
that the beam is run in the neutrino mode 
first and the true hierarchy turns out to be normal and 
$\stcht$ not too small. 
In that case,
we would see large number of events in the detector and 
know without ambiguity both the mass hierarchy as well as the 
value of $\theta_{13}$. However,
if the experiment is run in say only,  
the neutrino mode and if we do not see 
enough number of events (as expected from large matter effects), it 
would be difficult to tell if the non-observation 
should be attributed to 
the true hierarchy being inverted or whether the 
mixing angle $\theta_{13}$ is very small. We would find ourselves 
in a similar impasse if a $\anue$ \bb is used and one 
observed a very small number of events. 
This deadlock would then be broken only 
if we run the \bb in both the neutrino and the antineutrino 
mode. 

Recall that a \bb facility  has the useful
feature that it can be operated 
simultaneously in the $\nue$ as well as $\anue$ mode, since 
both the $^8B$ and $^8Li$ ions are positively charged, and 
therefore move in the same direction in the storage ring. 
The only constraint is that, since the magnetic field would 
be the same for both channels, the different charge-to-mass
ratio for the two candidate ions would fix the boost for 
the antineutrino channel to be 0.6 times that in the 
neutrino channel. Therefore, we could work within a
set-up where we have simultaneous $\nue$ and $\anue$ beams 
with say $\gamma=500$ and 300 respectively for the two channels. 
Since INO-ICAL would have excellent charge identification capability, 
it would distinguish the $\mu^-$ produced from oscillated $\numu$ 
and $\mu^+$ produced from oscillated $\anumu$, 
when the $\nue$ and $\anue$ beams are run simultaneously. 
In this scenario, unless $\theta_{13}$ is extremely small, we 
would observe an enhancement in the number of events due to 
earth matter effects in one of the channels, irrespective of 
the true mass hierarchy. Therefore, a direct comparison of the
number of $\mu^-$ and $\mu^+$ events would immediately tell us 
the mass hierarchy for $\theta_{13}$ not small. If 
the true value of $\theta_{13}$ is indeed chosen to be very small
by Nature, then we would see no enhancement in the event rate in 
either the neutrino or the antineutrino channel and one would 
be able to place an upper limit on the value of $\stch$ unambiguously. 

We have checked that with $\nue$ and $\anue$ beams running simultaneously 
with $\gamma=500$ and 300 respectively for the two channels, 
we could obtain a $\theta_{13}$ sensitivity of 
$\stch<4.5\times 10^{-3}$ at \sig,
irrespective of the true mass hierarchy, after 10 years of running.
We would have sensitivity to the neutrino mass hierarchy for at least
$\stch>2.1\times 10^{-2}$ at \sig{} after 10 years of simultaneous running, 
again irrespective of whether the true 
hierarchy was normal or inverted. One could argue that running the 
beam successively with $\gamma=500$, first in the $\nue$ 
mode followed by the $\anue$ mode would give an unambiguous $\theta_{13}$
sensitivity limit of $\stch < 1.5\times 10^{-3}$ at \sig{} and 
the mass hierarchy sensitivity limit of $\stch>8.2\times 10^{-3}$
at \sig. However, note that it is not possible to run the 
$\anue$ beam with the same $\gamma$ as the $\nue$ beam without 
any accelerator upgrade. Running the \bb simultaneously in both 
polarities has the additional advantage that if $\stcht$ is indeed 
large in Nature and within reach of the CERN-INO \bb set-up, 
then one would have a surplus of either $\mu^-$ or $\mu^+$ events 
in INO-ICAL. This would directly tell us the neutrino mass hierarchy
unambiguously.

\section{Conclusions}

We explored the physics potential of an experimental program with the 
\bb source ($\nu_e$ or $\bar{\nu}_e$) at CERN and a 
50 kton magnetized iron detector at INO. The 
CERN-INO distance is very close to the magic baseline, which is 
ideal for performing a degeneracy-free measurement of the mixing 
angle $\theta_{13}$ and the neutrino mass hierarchy. The 
large baseline also opens up the 
possibility for large matter effects. For the CERN-INO 
baseline, the
largest earth effects corresponding to resonant conversion 
of (anti)neutrinos in matter, happen for (anti)neutrino energy of
$E\simeq 6$ GeV.  $\nu_\mu$ or $\bar{\nu}_\mu$ produced at these
energies through oscillations can be 
effectively detected in magnetized iron calorimeters {\em via}
the $\mu^\mp$ they produce, the    
detection energy threshold being about 1 GeV. 
For the most common ions used in \bb studies, this range
of (anti)neutrino energies can be achieved for Lorentz factor 
$\gamma \gtap 10^3$. In this paper we have considered the 
new possibility of using the $^8B$ and $^8Li$ as source ions for
$\nue$ and $\anue$ \bb respectively. Owing to the large end-point 
energy of these ions, the resultant energy spectrum of the 
Lorentz boosted \bb for these ions 
is harder by more than a factor of 3, compared to the standard 
alternative ions $^{18}Ne$ and $^6He$. Therefore, with these ions 
we would get a \bb flux peaked at about 6 GeV, for plausible
values of the Lorentz boost $\gamma$ in the range 250-500.  
We showed that for this \bb flux from CERN, 
it will be possible to get 
an essentially background free
measurement of the ``golden channel'' probability $P_{e\mu}$,
through observation of $\mu^-$ (or $\mu^+$)
events in the ICAL detector at INO. 
We argued that the number of events in this experimental set-up is 
dictated directly by the extent of 
near-resonant earth matter effect, which 
significantly enhances the transition probability $P_{e\mu}$. 
The extent of earth matter effect in turn is governed by the 
neutrino mass hierarchy and the value of $\stch$. The 
CERN-INO \bb experiment that we propose 
therefore emerges as a powerful tool to 
pin down the neutrino mass hierarchy and $\theta_{13}$.

We simulated the prospective data in this experimental set-up 
for 5 years and 10 years of running of the \bb in either the 
neutrino or the antineutrino mode and presented the 
sensitivity results from a rigorous $\chi^2$ analysis, 
after marginalizing 
over all the oscillation parameters and taking into account
the systematic uncertainties coming from 
the \bb source, the detector, as well as the theoretical 
calculation of the interaction cross sections. For $\gamma$=500, if true 
value of $\theta_{13}$ was zero, this experiment with 60\% 
detection efficiency and 5 years of running would put a 
limit of $\stch < 2.1 \times 10^{-3}$ ($\stch < 1.5 \times 10^{-3}$)
at \sig{} C.L. 
with the $\nue$ ($\anue$) \bb for the normal (inverted) 
mass hierarchy. With 80\% detection efficiency and 10 years 
of running the corresponding limits would be 
$\stch < 8.4 \times 10^{-4}$ ($\stch < 5.7 \times 10^{-4}$)
at \sig{} C.L. 
with the $\nue$ ($\anue$) \bb for the normal (inverted) 
mass hierarchy. 
With 60\% detection efficiency and 5 years of running 
the wrong hierarchy could be ruled out at the \sig{} C.L. 
if $\stcht > 1.0 \times 10^{-2}$ ($\stcht > 8.2 \times 10^{-3}$)
with the $\nue$ ($\anue$) \bb for the true normal (inverted) 
mass hierarchy. If the detection efficiency was 80\% and 
running period 10 years the corresponding sensitivity would be 
enhanced to $\stcht > 8.5 \times 10^{-3}$ ($\stcht > 6.9 \times 10^{-3}$)
respectively at the \sig{} C.L. We showed that if $\stcht=0.05$, 
$\gamma=500$ and the true hierarchy was normal, then the 
inverted hierarchy could be disfavored at \sig{} C.L. with 
a $\nue$ \bb within just 9 months of running. 
Finally, we considered the 
scenario where the $\nue$ and $\anue$ flux could be produced 
simultaneously at the CERN \bb facility. Since ICAL at INO 
would be magnetized and hence would possess charge identification 
capability, we argued that an unambiguous information 
on $\theta_{13}$ and the mass hierarchy could be obtained 
within a much shorter time scale if $\stcht$ is large. If 
$\theta_{13}$ is very small, one would not 
be able to determine the 
mass hierarchy, but could 
still save time in 
putting an upper bound on the mixing angle $\theta_{13}$.

In conclusion, the sensitivity for $\stch$ and the neutrino 
mass hierarchy that is expected in the CERN-INO \bb experiment 
appears to be better than most of the other experimental proposals 
involving superbeams and \bb and can only be surpassed by 
the NuFact-II experiment.

\vskip 1.0cm
{\Large \bf Acknowledgements}
\vskip 0.3cm
\noindent
The authors acknowledge help from the INO collaboration and 
thank F. Terranova for a useful communication.
S.K.A. is grateful to Abhijit Samanta for discussions. S.C. wishes to 
thank Walter Winter, Andrea Donini and Pasquale Migliozzi for 
helpful discussions. S.C. acknowledges support from 
the University of Oxford and PPARC during the initial part
of this work.


\end{document}